\documentclass[trackchanges]{aastex701}
\usepackage{hyperref}
\usepackage{caption,subcaption}
\usepackage{tabularx}
\usepackage{amsmath}
\usepackage{physics}
\usepackage{tikz}
\usepackage{pgfplots}
\pgfplotsset{compat=1.18}
\usetikzlibrary{shapes,arrows,positioning,backgrounds,external}
\usepgfplotslibrary{patchplots,fillbetween}

\newcommand{\parens}[1]{\left(#1\right)}
\newcommand{\mat}[1]{\begin{bmatrix}#1\end{bmatrix}}

\begin{document}

\title{Control algorithms for dual-wavefront sensor single-conjugate adaptive optics}

\author[0000-0002-5669-035X]{Aditya R. Sengupta}
\affiliation{Department of Astronomy and Astrophysics, University of California, Santa Cruz, CA 95064, USA}
\email[show]{adityars@ucsc.edu}

\author[0000-0002-4300-3426]{Lisa A. Poyneer} 
\affiliation{Lawrence Livermore National Laboratory, 7000 East Avenue, Livermore, CA 94550, USA}
\email{}

\author[0000-0003-3978-9195]{Benjamin L. Gerard} 
\affiliation{Lawrence Livermore National Laboratory, 7000 East Avenue, Livermore, CA 94550, USA}
\email{}

\author[0000-0003-0054-2953]{Rebecca Jensen-Clem} 
\affiliation{Department of Astronomy and Astrophysics, University of California, Santa Cruz, CA 95064, USA}
\email{}

\author[0009-0002-4340-0496]{Aaron J. Lemmer} 
\affiliation{Lawrence Livermore National Laboratory, 7000 East Avenue, Livermore, CA 94550, USA}
\email{}

\begin{abstract}

High-contrast imaging systems using active control with adaptive optics (AO) are often limited by non-common path (NCP) aberrations that are seen only at the final science image. AO systems employing focal-plane wavefront sensors (FP-WFSs) are able to simultaneously correct NCP aberrations and measure science images, but they typically require a second stage of control that adds system cost and complexity. We present control algorithms to augment AO systems with FP-WFSs within their existing control setup. We demonstrate inter-arm NCP aberration transfer can be mitigated through temporal filtering, present frequency- and time-domain validation of controller stability and performance, and discuss the optimality of the chosen controllers. This work will enable the development, testing, and installation of FP-WFS technologies for direct imaging of exoplanets.

\end{abstract}



\section{Introduction}
\label{sec:intro}

Adaptive optics (AO) systems on ground-based observatories have enabled the direct imaging of extrasolar planets at high contrast and precision \citep{scexao}. AO systems involve real-time control of optical aberrations, which are sensed by a wavefront sensor (WFS) whose readings are used to compute and send commands to a deformable mirror (DM). These aberrations are principally induced by the atmosphere, but also arise from thermal and mechanical effects. In the extreme (high-contrast) AO case, non-atmospheric aberrations can constitute a significant part of the AO error budget (e.g., \citealt{Martinez2012}). In particular, we observe \textit{non-common-path} (NCP) aberrations that are induced only on the science imaging arm and not seen by the WFS.

Focal-plane wavefront sensing (FP-WFS) technologies intended to correct NCP aberrations have been demonstrated and deployed on sky \citep{Haffert2022, Males2024, Lin2023}. These apply a transformation to light at the plane of the imaging camera in order to sense aberrations directly on final science images. Examples include the self-coherent camera (SCC) \citep{Gerard2018} and the photonic lantern (PL) \citep{Norris2020}.

FP-WFSs are therefore useful to reach the high precision required by extreme AO, but are limited relative to traditional pupil-plane WFSs. They usually require longer integration times due to their dual role as science imagers, and they are often limited in the amplitude or spatial order of the aberrations they can sense. For instance, the PL can only sense out to a spatial order corresponding to the physical size of its input; for a 19-port lantern, which has been tested as a wavefront sensor on sky \citep{Lin2023} and in the lab \citep{Norris2020, Sengupta2024}, aberrations past 4 $\lambda/D$ create PSF features that fall outside the input and therefore cannot be seen. This means high-order aberrations can cause poor initial coupling that the PL alone is unable to correct. Relatedly, \cite{Gerard2021} showed for the SCC that the linear and capture range are around 0.8 and 1.9 radians rms, respectively. Focal-plane WFSs are thus generally required to operate jointly with pupil-plane WFSs.

Existing extreme AO systems, such as MagAO-X \citep{Males2024}, run FP-WFS as a second-stage system. This involves installing a FP-WFS and a dedicated DM instead of a science imaging camera, downstream of a pupil-plane WFS and DM, to make a multi-stage AO system. Second-stage systems require running a separate control loop, and also require procuring and managing an extra DM; these increase system cost and complexity. A completely software-based solution for FP-WFS testing that does not require a dedicated DM would expand the potential scope of technology demonstrations and allow for easier deployment of these instruments to existing observatories. FP-WFS capabilities within a single-conjugate AO system would therefore enable more rapid development of improved high-contrast imaging methods. Future space-based exoplanet imaging systems will deploy a low-order WFS in addition to the coronagraphic image/WFS (e.g., the Roman Coronagraph Instrument \citep{PoberezhskiyRomanCGI}), and phasing errors on segmented primary mirrors may also be a source of time-varying aberrations that would be sensed at a point other than the science imaging plane \citep{Douglas2019}. In both cases, temporal NCP errors differing between the two WFS arms may similarly limit performance goals.

Static NCP errors (which do not vary with time) can be corrected using one DM that is common-path to both WFSs by altering the reference on the pupil-plane WFS such that it produces the desired wavefront at the focal plane (e.g. \citealt{Vigan2018}). We refer to this as `setpoint control'. This motivates a control algorithm that could extend this to dynamic NCP errors. Existing focal-plane wavefront control techniques, such as the Fast \& Furious \citep{ff14} and DrWHO \citep{Skaf2022} algorithms, are able to correct dynamic NCP errors, but have not yet been studied at loop speeds where correction of residual atmospheric turbulence becomes relevant as is proposed here. {For instance, the on-sky demonstration of DrWHO updated the reference image at a rate of 0.03 Hz, and therefore did not control higher-speed (1-100 Hz) atmospheric residuals.} 

{Setpoint control at high speeds, in which the reference for the pupil-plane WFS is altered every frame or at frequencies above $\sim$100 Hz, would induce new controller dynamics. For instance, a stable high-speed setpoint controller would likely require multiplying the reference by a gain to maintain loop stability. Optimizing controller performance while maintaining stability margins would require deriving a new control framework, consisting of transfer functions from each disturbance to each WFS readout that model the setpoint-updating operation. Such a framework has not yet been developed and would be of interest for future work. We consider only the case where signals from both WFSs are used as input to separate controllers, as it is simpler to implement using largely the same calibration and control steps as the single-WFS control case.}

\begin{figure}
    \includegraphics[width=\linewidth]{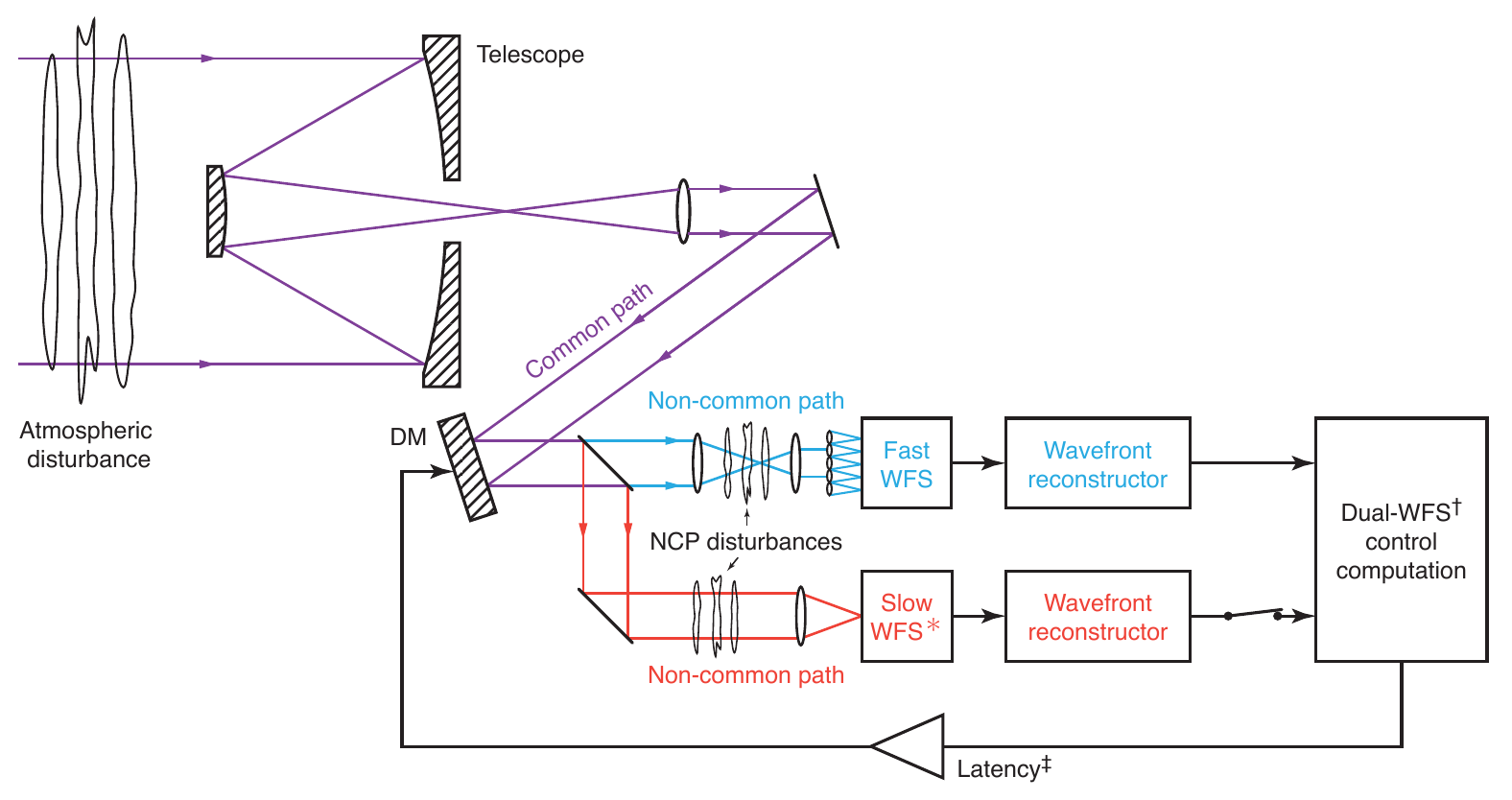}
    \caption{The dual-WFS single-DM control scheme for a representative astronomical AO system, indicating the introduction of common- and non-common-path disturbances. The common optical path is shown in purple and the non-common optical paths are shown in blue and red. NCP disturbances are shown in localized points in each beam for visualization purposes, but can in general arise anywhere along the non-common paths. \textsuperscript{*}The slow WFS is also the science imager. \textsuperscript{\dag}If only the fast WFS is used for feedback, this block represents the standard single-WFS control computation and corresponds to the schematic switch shown in the open state. \textsuperscript{\ddag}Here latency represents all delays, including servo, processing, network/propagation, and mechancial latency of the DM.}
    \label{fig:boxdiagram}
\end{figure}

In this work, we present control algorithms that combine readings from two wavefront sensors, operating at different frame rates and with time-varying NCP aberrations on both, to produce DM commands that minimize the error seen by the focal-plane WFS while remaining within the dynamic range of the pupil-plane WFS. We also develop a controls framework for this dual-WFS single-conjugate AO problem. Figure~\ref{fig:boxdiagram} shows a representative diagram of the control configuration considered in this work.

Due to the NCP aberrations seen by both WFSs, the usual control scheme using an integrator would lead to inter-arm signal transfer, in which the DM moves to correct NCP aberration on one arm, thus degrading wavefront quality on the other. Previous work on an analogous problem involving two-DM control \citep{Gavel2014} resolved this issue by temporally filtering the signal seen by the faster DM in order to suppress modes within the control space of both DMs. A control algorithm in which integrator control commands from two WFSs are {temporally} filtered to avoid frequency-domain overlap has shown stable closed-loop performance and correction of NCP aberrations in simulation \citep{Sengupta2024} and in the lab \citep{Gerard2023}. A particular motivation of this work was to develop control algorithms that could rigorously impose constraints on the error seen at a particular WFS (e.g. ensuring the total error seen by an SCC is always less than 0.8 rad rms, in order to remain within its linear range). This motivated our choice to focus on separate controllers with temporal filtering to explicitly handle NCP transfer. The problem formulation that we present may additionally be relevant to enable development of high-speed setpoint control algorithms.

In preparation for on-sky technology demonstrations of FP-WFSs, we extend this architecture by optimizing the controller for a range of NCP aberration and WFS noise strengths, and explore the space of controllers in order to achieve optimal performance for this control problem. Ultimately, the motivation for dual-WFS control is to allow a second WFS to enable lower closed-loop wavefront error than a single-WFS AO system could reach, such as improved sensitivity to measurement noise and/or aliasing. However, these benefits have been already studied in other papers \citep{Gerard2021,Chambouleyron2023}, and so in this paper we will consider the bandwidth error and measurement noise implications to the dual-WFS control problem from a WFS-agnostic approach (i.e., assuming the two WFSs are each the same type of WFS). 

The remainder of this paper is structured as follows. Section~\ref{sec:problem} describes the models used for the AO system, disturbance processes, and control strategies considered in this work. Section~\ref{sec:freq} describes the results of optimization over each control strategy in the frequency domain. Section~\ref{sec:time} describes the evaluation of the resulting optimal controllers in the time domain. Section~\ref{sec:further_work} discusses prospects for further improving the control strategies considered in this work. Section~\ref{sec:conclusion} is a conclusion.

\section{Problem formulation}\label{sec:problem}
\subsection{Signal flow}
We consider an AO system in which incoming starlight, from which the relevant signal is the atmospheric aberration $\phi$, first encounters a DM whose shape is set in closed loop. The resulting error signal $\epsilon$ is observed along two paths, corresponding to the slow and fast WFS, combined with NCP aberration signals $L_\text{slow}$ and $L_\text{fast}$. Note that in our notation the subscripts `slow' and `fast' refer to the disturbance signals seen by the slow and fast WFS, and do not suggest that these signals have distinct temporal properties from one another. We consider signals $X$ and $Y$, which represent the wavefront as seen by the slow (focal-plane) and fast (pupil-plane) WFS, respectively:

\begin{align}
    X &= L_\text{slow} + \epsilon\\
    Y &= L_\text{fast} + \epsilon.
\end{align}

We observe $X$ and $Y$ after delays due to the respective WFS readouts, and with the addition of WFS noise terms $N_\text{slow}$ and $N_\text{fast}$. In this work we consider a ratio of fast to slow frame rates of $R = 10$. The resulting signals are processed by the respective controllers $C_\text{slow}$ every $R$ timesteps and $C_\text{fast}$ every timestep. The control commands are added together, after computational delays on both that are possibly different (we take them both equal to 1 millisecond). This combined control command has a zero-order hold due to maintaining the previous DM command until the new one can be applied. The control computation and effect therefore incorporates delays from the end of the WFS exposure to the end of the DM facesheet settling time, including camera readout, latency, and physical DM response.

\begin{figure}
    \centering
    \includegraphics[width=\linewidth]{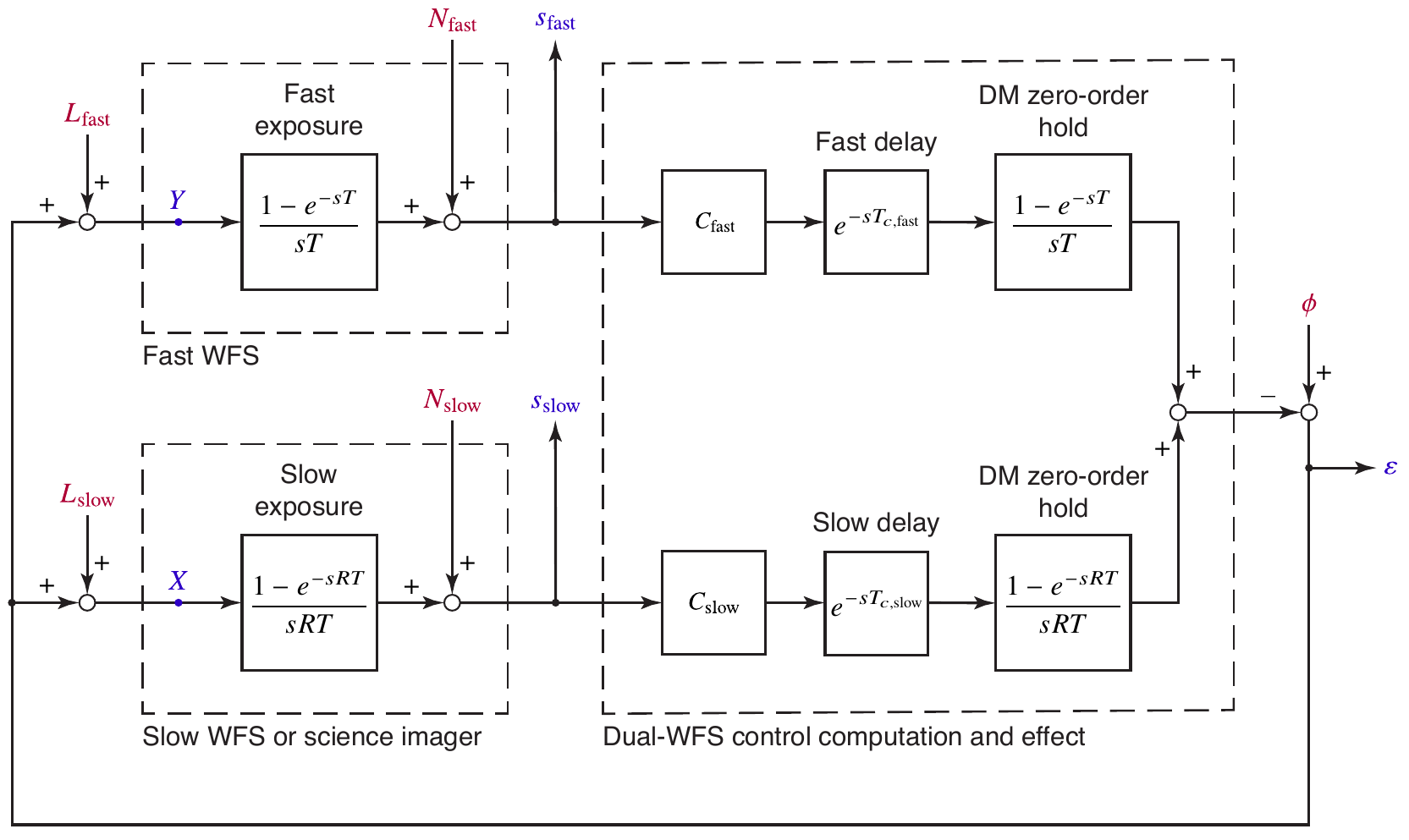}
    \caption{A block diagram of two-WFS control, indicating the different frame rates for the two WFSs, to enable controller analysis and optimization. Disturbances are shown in red and relevant intermediate and output signals are shown in purple.}
    \label{fig:blockdiagram}
\end{figure}

Figure~\ref{fig:blockdiagram} shows a schematic of this AO system. We model the control system in the frequency domain and in continuous time using the Laplace variable $s = 2\pi i f T$. We use a plant model incorporating the WFS exposure time, computational delay, and zero-order hold from \cite{veran2009type}.

We consider a fast WFS exposure time interval of $T = 0.001$s and a slow time interval of $RT = 0.01$s. For simplicity, as we introduce this topic from an AO control perspective, Fig.~\ref{fig:blockdiagram} enables an analytical model of the bandwidth and measurement error terms of an AO error budget. Additional terms will be considered in 2D simulations later in \S\ref{sec:2dsims}.

\subsection{Closed-loop error metric}
We optimize controllers to minimize the closed-loop error at $X$ while keeping the closed-loop error at $Y$ below a certain threshold, which we set based on the dynamic range of the focal-plane WFS, motivated above in \S\ref{sec:intro} and generally kept to $< 1$ rad rms throughout this paper. We compute transfer functions from the five disturbance inputs ($\phi$, $L_\text{fast}$, $L_\text{slow}$, $N_\text{fast}$, $N_\text{slow}$) to $X$ and $Y$, which we denote $X/\phi$, $X/{L_\text{fast}}$ and similarly for the others. Derivations for these transfer functions are provided in Appendix A. We then compute a closed-loop error in the frequency domain. We define the `integrand' error terms respectively for the atmosphere, NCP, and noise components as
\begin{align}
    IE_{X,\text{atm}}(f) &= p_\text{atm} (f) \abs{\frac{X}{\phi}}^2 \label{eq:ie_atm}\\
    IE_{X,\text{NCP}}(f) &= p_\text{NCP} (f) \qty(\abs{\frac{X}{L_\text{fast}}}^2 + \abs{\frac{X}{L_\text{slow}}}^2) \label{eq:ie_ncp}\\
    IE_{X,\text{noise}}(f) &= p_\text{noise} (f) \qty(\abs{\frac{X}{N_\text{fast}}}^2 + \abs{\frac{X}{N_\text{slow}}}^2)\label{eq:ie_noise},
\end{align}
where $p$ denotes the open-loop power spectral density (PSD) due to each component. We assume $p_\text{NCP}$ and $p_\text{noise}$ are the same for the two sensors throughout this work, but this may be different in the general case. { We assume all disturbance sources are temporally uncorrelated, allowing us to add their power spectra in quadrature. We note that actual systems may have correlations in the NCP aberrations on both WFS arms, which we neglect in the interest of generality. Considering such correlations would require modeling a particular optical design and the potential sources of temporal NCP disturbances and then adjusting the error metric accordingly.}

We consider control over a relevant frequency range $[f_{\min}, f_{\max}]$. We take $f_{\min} = 1$ Hz since we later consider time-domain simulations of length $1$s. We take $f_{\max} = 500$ Hz, equal to the Nyquist limit relative to the fast sensor. For $N_{\text{slow}}$ the Nyquist limit is $50$ Hz, so we consider $|X/N_{\text{slow}}|$ and $|Y/N_{\text{slow}}|$ to be zero for $50 \text{Hz} \leq f \leq 500$ Hz.

We compute the closed-loop error by summing the error terms in Equations~\ref{eq:ie_atm}-\ref{eq:ie_noise} together, integrating over the chosen frequency range, and taking a square root:
\begin{align}
    \text{Error at } X = \sqrt{\int_{f_{\min}}^{f_{\max}} \mathrm{d}f \left(IE_{X,\text{atm}}(f) + IE_{X,\text{NCP}}(f) + IE_{X,\text{noise}}(f)\right)}.
\end{align}

The analogous expressions apply for $Y$. Since the atmosphere is common-path and the noise terms only impact the signal seen going into the WFSs due to the feedback, the atmosphere and noise error terms (Equations~\ref{eq:ie_atm} and \ref{eq:ie_noise}) are the same for $X$ and $Y$, and are only different for NCP aberrations.

\subsection{Open-loop power spectrum models}
This objective function requires a model for the open-loop PSDs due to each disturbance. We assume a von K\'{a}rm\'{a}n profile for atmosphere and NCP with different $r_0$ and wind speeds. This is a standard model for atmospheric aberrations \citep{Tokovinin2007}, and is also employed here for NCP aberrations to reflect the impact of air-based turbulence within an optical bench. The PSD is
\begin{align}
    p_\text{VK}(f) = 0.25 \parens{r_0 / \text{m}}^{-5/3} \parens{f + f_0}^{-2} \text{rad}^2/\text{Hz},
\end{align} 
where $r_0$ is the Fried parameter and $f_0$ is a characteristic frequency set to match real power spectra. We take $r_0 = 0.1031$ m (i.e., $\sim$1" seeing at $\lambda$ = 500 nm) and $f_0 = 1$ Hz for the atmosphere (i.e., a 10 m/s frozen flow ground layer for a $D$ = 3 m telescope), and $f_0 = 0.001$ Hz for NCP. { Since $f_0$ for the NCP power spectrum is significantly lower than for the atmosphere, the open-loop power spectrum for NCP aberrations resembles a power-law at the $>1$ Hz frequencies considered here. Using the von K\'{a}rm\'{a}n power spectrum allows us to interpret the size scale of NCP turbulence in terms of $r_0$, as we would with atmospheric aberrations.} We will consider a range of values for $r_0$ for the NCP power spectrum, between 0.6m and 6.0m, which correspond to spatial wavefront errors with 140 nm rms and 21 nm rms, respectively. This upper limit on NCP $r_0$ is empirically motivated, considering that operations of AO systems have shown NCP strength is likely below the level of AO residuals.

We take WFS noise to be a white noise process (i.e., its PSD is a constant) whose strength is set by a noise/atmosphere crossover frequency $f_\text{cross}$. We set the strength of noise at all frequencies to be equal to the strength of the atmosphere at $f_\text{cross}$.

\begin{figure}
    \centering
    \includegraphics[width=\textwidth]{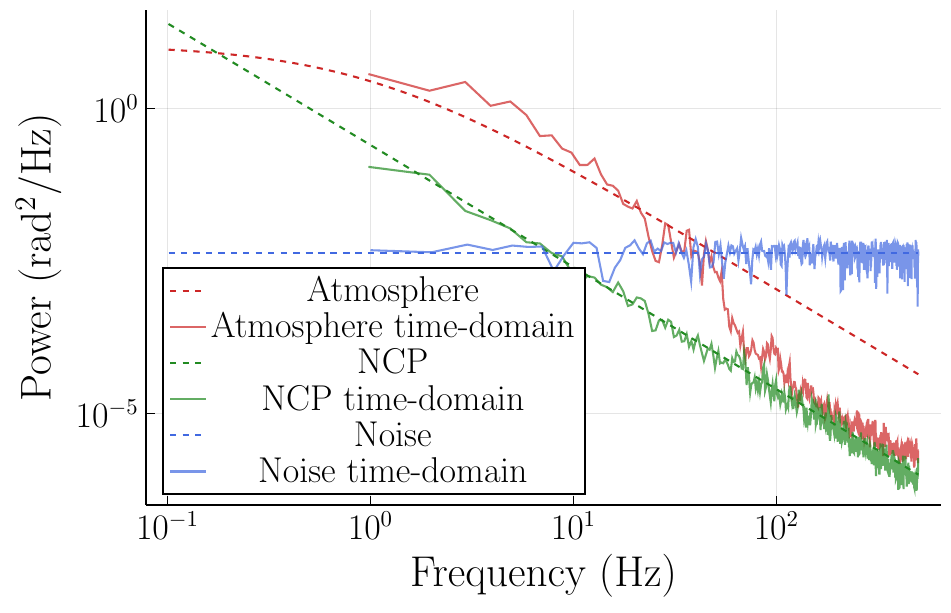}
    \caption{The estimated open-loop PSDs for the atmosphere, non-common-path aberrations, and noise as seen in the time-domain AO simulation (\S\ref{sec:2dsims}). We take $r_0 = 1$ m for NCP aberrations and $f_\text{cross} = 50$ Hz. The dashed lines show the analytic power spectrum models considered for frequency-domain optimization, and the solid lines show the PSDs of a representative 1-second time series.}
    \label{fig:openloop_psds}
\end{figure}

Figure~\ref{fig:openloop_psds} shows representative open-loop PSDs considered in this work. For optimization we use the analytic (dashed) profiles. { The overall frequency-domain model was validated by comparing 1D time- and frequency-domain realizations of von K\'{a}rm\'{a}n turbulence, both in open loop and closed loop.} In time-domain simulations, we generate NCP and noise time-series such that their estimated power spectra closely match the analytic ones, by taking an inverse Fourier transform of the analytic power spectra and augmenting the result with random phases. The atmospheric time-series are generated as detailed in \S\ref{sec:2dsims}. These deviate from the von K\'{a}rm\'{a}n profile at high frequencies; this difference may skew optimization results. We attribute this to the frequency-domain model not correctly modeling the average actuator temporal PSD { from the later time-domain simulations}. Modeling 2D frozen flow in the frequency domain with a single power law and turnover frequency is representative for low order Zernike modes such as tip/tilt \citep{Noll1976}. This does not hold for higher order Fourier modes \citep{Poyneer2009}, in which case the average actuator PSD would be more realistic. However, we found that this 1D approximation was sufficiently representative of later 2D time-domain simulation results (\S\ref{sec:2dsims}). { We} therefore decided not to develop a more complex higher order analytical model more realistic of the average actuator PSD. { In closed loop, we compare 1D frequency-domain and time-domain power spectra of control residuals and find exact agreement over the frequencies covered by the time horizon of the time-domain simulations.} We note that controller optimization does not depend on an analytic open-loop power spectrum model, and in practical settings it can be carried out with respect to empirically estimated power spectra.

\subsection{Modeling the slow sensor via sample-and-hold}\label{sec:sampleandhold}

The dual-WFS problem cannot be perfectly modeled within the frequency domain because such a description assumes that the system is time-invariant; that is, all transfer functions are independent of the particular timestep and can be delayed for an arbitrary amount of time. This does not apply to the two distinct frame rates in the dual-WFS problem, since the slow WFS will read out every $R$ steps and will not read out in the intermediate steps. Therefore, our frequency-domain model assumes that we read out a moving average of the previous $R$ frames of the slow WFS signal every timestep. This approximation mostly affects performance above the Nyquist cutoff for the slow WFS, which in this work is 50 Hz. Since the aberrations under consideration have most of their power at lower frequencies than this, we do not expect simulation and optimization results to be significantly impacted by this difference. In Section~\ref{sec:time}, we validate that the closed-loop errors predicted under various controllers by the frequency-domain and time-domain model are substantially similar. 

\subsection{Control strategy formulations}

We consider four control strategies:

\begin{enumerate}
    \item Integrator on one WFS (`Single IC')
    \item Integrator on two WFSs (`Double IC')
    \item Integrator on two WFSs, with an autoregressive order-1 (AR(1)) high-pass filter on the fast WFS outputs (`Double IC-HPF')
    \item Linear-quadratic Gaussian (LQG) controller for an AR(1) process on two WFSs, with an AR(1) high-pass filter on the fast WFS outputs (`Double LQG-IC-HPF')
\end{enumerate}

We provide models for these controllers in the Laplace domain. 

\subsubsection{Integrator}
For the integrator with gain $g$ and leak $l$ (fixed to 0.999 throughout this work), we have
\begin{align}
    C_\text{integrator}(s) = \frac{g}{1 - l \exp(-T_s s)},
\end{align}
where $T_s = T = 0.001$s on the fast WFS and $T_s = RT = 0.01$s on the slow WFS. 

\subsubsection{Filter}
We consider AR(1) high-pass filters whose time-domain behavior is defined by the following relationship between an input signal $x$ and an output signal $y$, as defined by \cite{Gavel2014}:
\begin{align}
    y[n] = \alpha y[n-1] + \alpha (x[n] - x[n-1]) \label{eq:filter_timedomain}.
\end{align}

These are parameterized by a coefficient $\alpha$. We optimize over the cutoff frequency $f_c$ of the filter, which is related to $\alpha$ according to
\begin{align}
    \alpha = \exp\left(-2\pi f_c T\right).
\end{align}

The transfer function for this filter is
\begin{align}
    C_\text{filter}(s) = \alpha \left(\frac{1 - \exp(-s)}{1 - \alpha \exp(-s)}\right).
\end{align}


\subsubsection{LQG integrator}
The optimal LQG controller for an AR(1) input process { has an integral action term and shows} improved bandwidth over the simple integrator. { This makes} it potentially able to deliver better performance and { presents} a useful test case to explore the space of controller designs. { We refer to these as LQG integrators (or `LQG-IC' in shorthand) as their parameters are set via empirical tuning, as with a regular integrator, rather than by modeling the system in state space, as is otherwise normal for LQG. These should be thought of as LQG models for a first-order turbulence model.} We follow the formulation of \cite{Looze}, as employed by \citealt{LLAMAS}. Our LQG integrator uses the following matrix as its evolution model, going into equation 18 of Poyneer \textit{et al.} 2023:
\begin{align}
    A = \mat{\alpha_\text{LQG} & 0 \\ 1 & 0},
\end{align}
where $\alpha_\text{LQG}$ is a free parameter whose role is analogous to that of the leak of the simple integrator. We also take the process noise term $\sigma_v^2$, which modifies the controller via the Kalman gain (Equations 22 and 23 of Poyneer \textit{et al.} 2023), as a free parameter.

For the slow LQG integrator, because the 1 ms delay is a $1/R$ frame delay for the slow controller, we modify the measurement model so that it observes a weighted average of the last two frames, weighted by a fraction $\texttt{computation\_delay}/R$ to the last frame and $1 - \texttt{computation\_delay}/R$ to the frame before that. (See Eq. 11 of Poyneer \textit{et al.} 2023).

\subsection{Stability and performance optimization}

We enforce stability by requiring a gain margin of 2.5 and a phase margin of $45^\circ$, matching that set by \citealt{veran2009type} and used by Poyneer \textit{et al.} 2023. We compute these using the plant transfer function, between the atmosphere and the DM feedback arm in open-loop. The plant transfer function is derived in the supplementary material. We identify the points on the Nyquist contour closest to $-1$ along the imaginary axis and the unit circle via root-finding.

We find optimal controllers for each case via an iterative grid search over all parameters. The grids are initially coarse and are refined around the solution to precisely resolve the optimal parameters.

\section{Frequency-domain optimization results}\label{sec:freq}

We visualize the stability, performance, and optimality of each controller. For each control strategy, we show the optimal controller for NCP $r_0 = 1$ m and $f_\text{cross} = 50$ Hz. We show:

\begin{itemize}
    \item the Nyquist diagram including gain and phase margins, to check stability;
    \item the error transfer functions (ETFs) from each disturbance source to $X$;
    \item the error transfer functions from each disturbance source to $Y$;
    \item the integrand error terms (Equations~\ref{eq:ie_atm}-\ref{eq:ie_noise}) for $X$ and the NCP integrand error term for $Y$;
    \item the open-loop power spectra (matching Figure~\ref{fig:openloop_psds}) due to each disturbance source, and the closed-loop power spectra at $X$ and $Y$; and
    \item the closed-loop error at $X$ as a function of each parameter being optimized over, to show the controller is locally optimal.
\end{itemize}

\subsection{Single integrator control}

The baseline strategy we consider is control using an integrator on the fast WFS, without making use of slow WFS measurements. We assume a leak of 0.999 and optimize over the gain. 

\begin{figure}
    \begin{subfigure}{0.33\textwidth}
    \includegraphics[width=\linewidth]{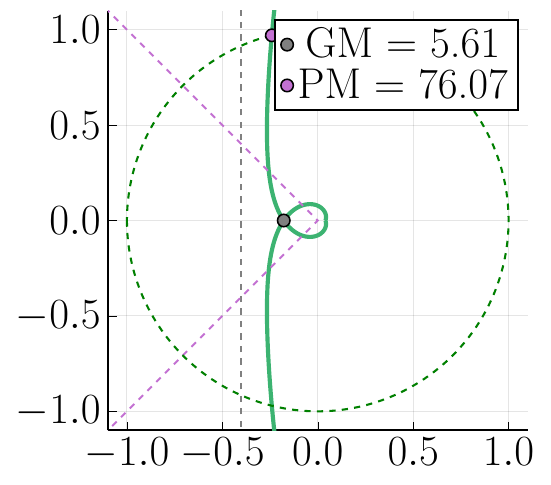}
    \caption{Nyquist diagram} \label{fig:nyquist_singlewfs}
    \end{subfigure}\hspace*{\fill}
    \begin{subfigure}{0.33\textwidth}
    \includegraphics[width=\linewidth]{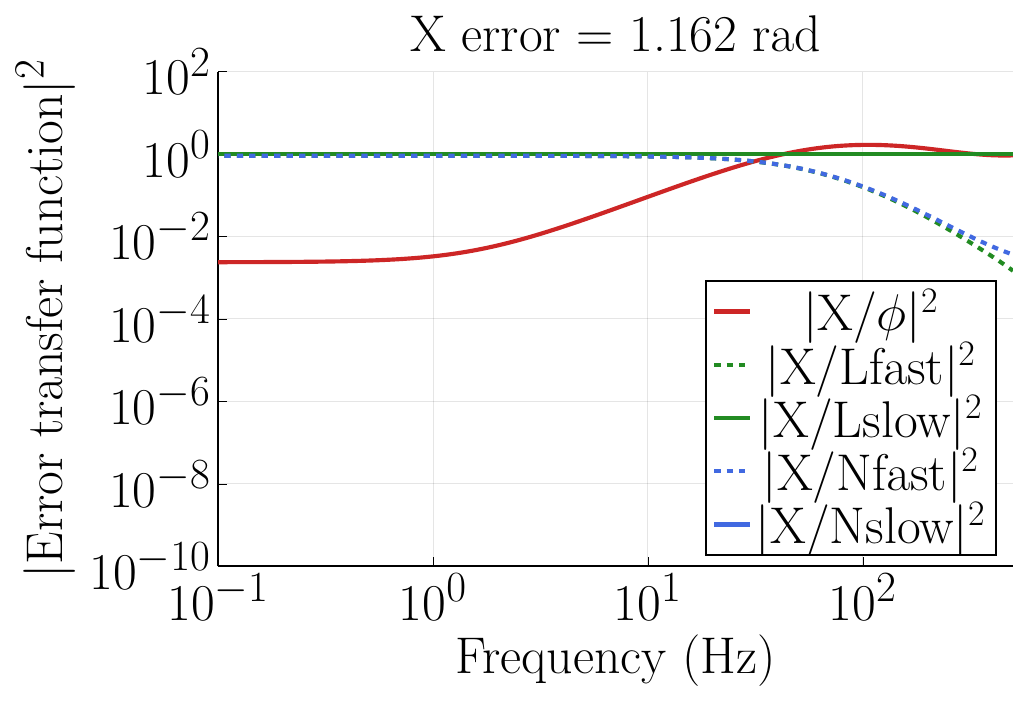}
    \caption{Error transfer functions to $X$} \label{fig:errx_singlewfs}
    \end{subfigure}\hspace*{\fill}
    \begin{subfigure}{0.33\textwidth}
    \includegraphics[width=\linewidth]{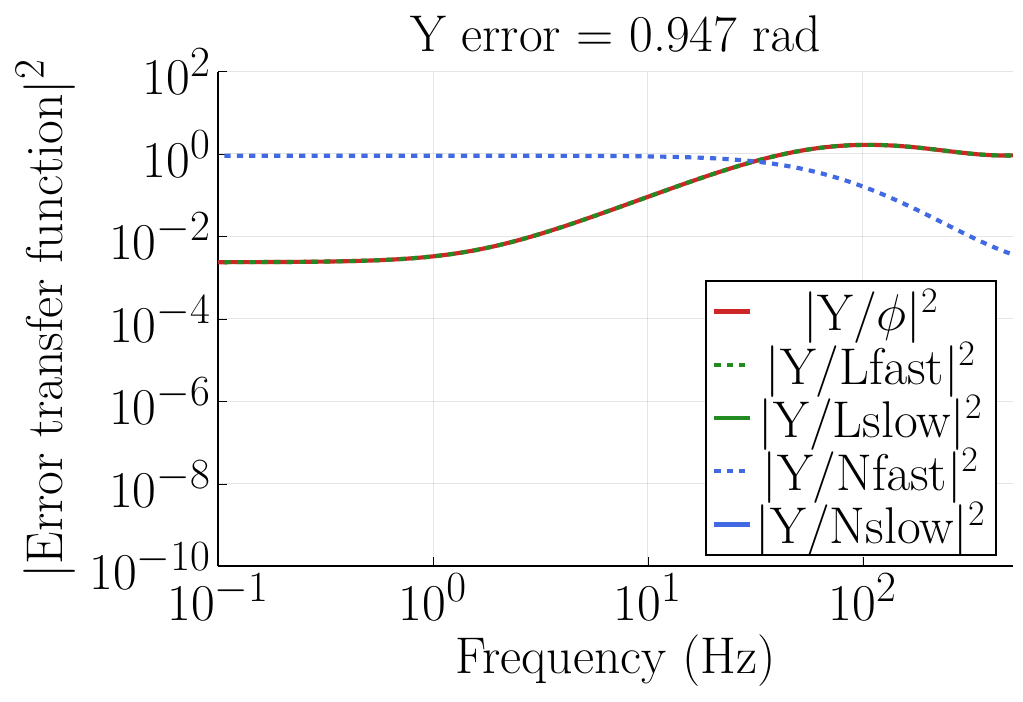}
    \caption{Error transfer functions to $Y$} \label{fig:erry_singlewfs}
    \end{subfigure}

    \medskip
    \begin{subfigure}{0.33\textwidth}
    \includegraphics[width=\linewidth]{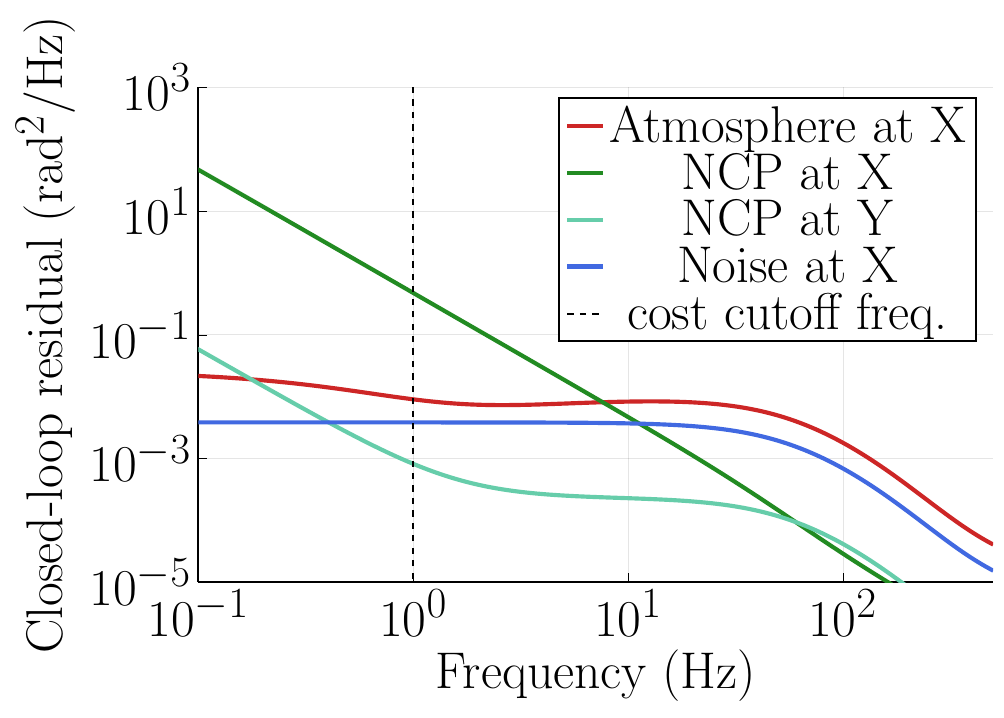}
    \caption{Closed-loop integrands} \label{fig:integrands_singlewfs}
    \end{subfigure}\hspace*{\fill}
    \begin{subfigure}{0.33\textwidth}
    \includegraphics[width=\linewidth]{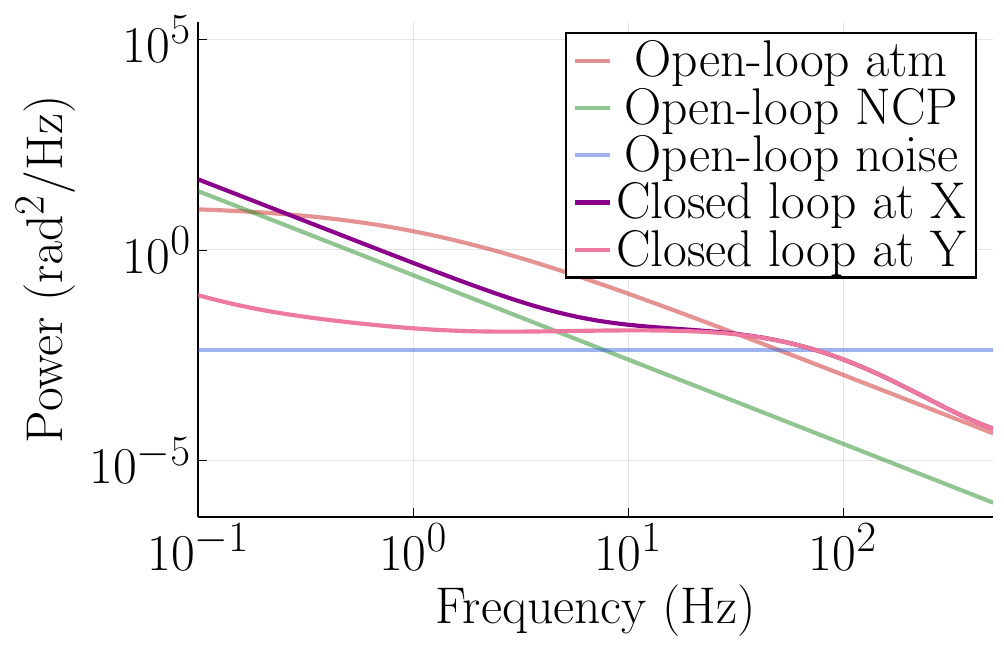}
    \caption{Open- and closed-loop PSDs} \label{fig:psds_singlewfs}
    \end{subfigure}\hspace*{\fill}
    \begin{subfigure}{0.33\textwidth}
    \includegraphics[width=\linewidth]{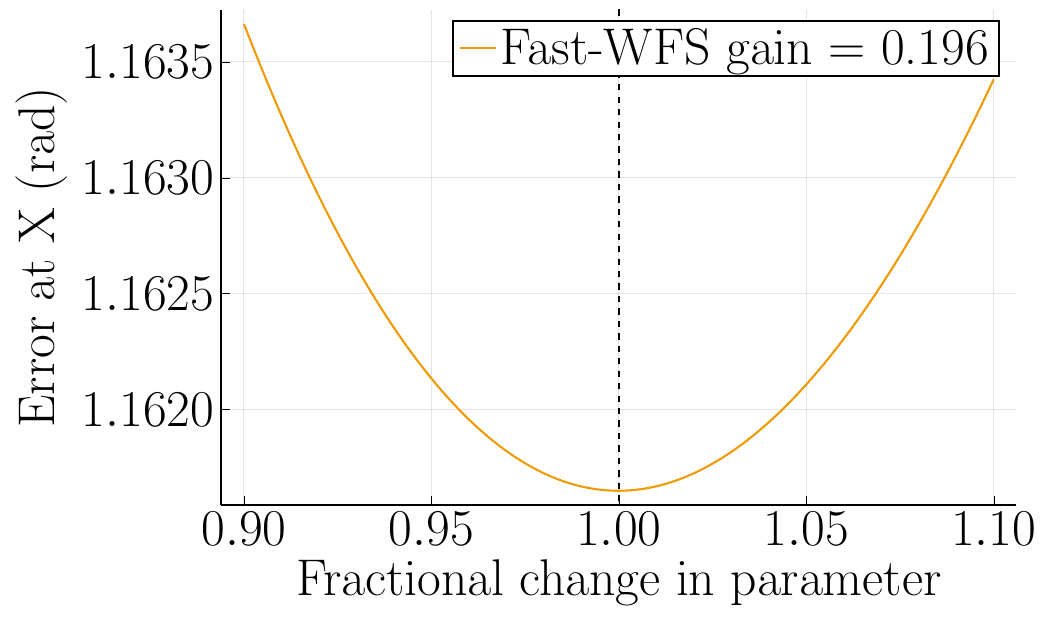}
    \caption{Error at $X$ as a function of gain} \label{fig:minplot_singlewfs}
    \end{subfigure}
    
    \caption{The single IC controller is unable to correct NCP aberrations seen only by the slow WFS. Since there is no slow WFS, there is no attenuation of $L_{\text{slow}}$ (solid green line in Figure~\ref{fig:errx_singlewfs}), and therefore NCP error at $X$ is a large part of the residual error (dark green line in Figure~\ref{fig:integrands_singlewfs}). Not attenuating slow NCP error results in a higher closed-loop error at $X$ than $Y$ (purple and pink lines in Figure~\ref{fig:psds_singlewfs}).} 
    \label{fig:singlewfs}
\end{figure}

Figure~\ref{fig:singlewfs} shows a representative optimal IC controller. In Figure~\ref{fig:errx_singlewfs}, we observe no attenuation of the slow NCP error at $X$ and no transfer of slow WFS noise to $X$ or $Y$, as expected due to the lack of slow WFS feedback. In Figure~\ref{fig:erry_singlewfs}, we also observe no transfer of the slow NCP error to $Y$, and we note that the error transfer function from $\phi$ and $L_\text{fast}$ to $Y$ are the same, since we do not control either term separately from the other.

Figure~\ref{fig:integrands_singlewfs} shows a significant contribution to the controller error due to not correcting the NCP aberrations at $X$. Together with Figure~\ref{fig:psds_singlewfs}, this indicates that better correction is needed particularly in the 1--50 Hz range, meaning significantly better performance is possible using the slow WFS measurements even at the slower frame rate.

\subsection{Double integrator control}

The double integrator has an opportunity to correct slow NCP aberrations. We optimize over the gain of both the fast and slow integrator.

\begin{figure}
    \begin{subfigure}{0.33\textwidth}
    \includegraphics[width=\linewidth]{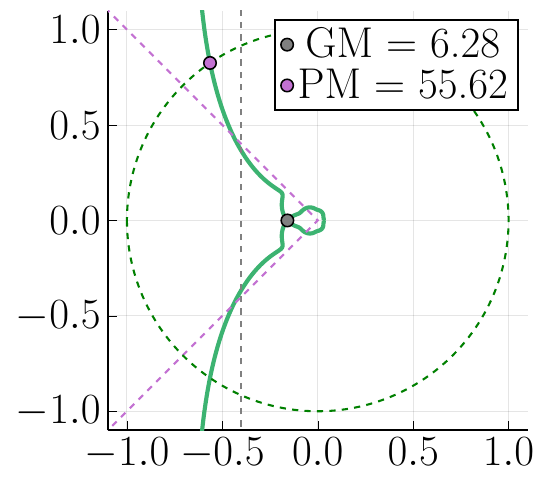}
    \caption{Nyquist diagram} \label{fig:nyquist_doublewfs}
    \end{subfigure}\hspace*{\fill}
    \begin{subfigure}{0.33\textwidth}
    \includegraphics[width=\linewidth]{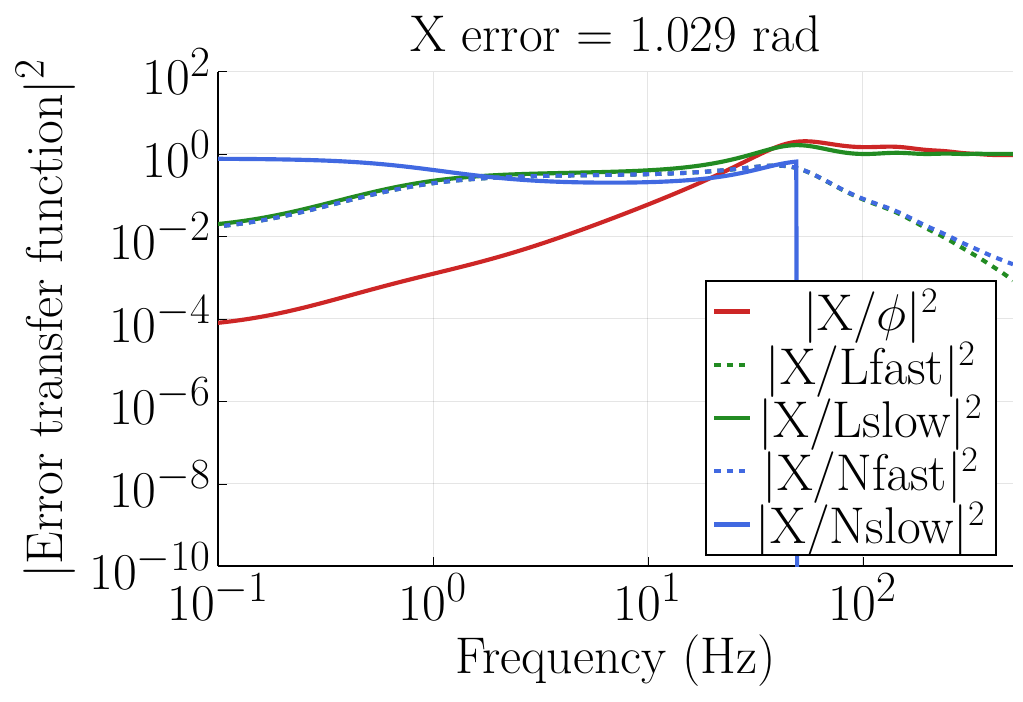}
    \caption{Error transfer functions to $X$} \label{fig:errx_doublewfs}
    \end{subfigure}\hspace*{\fill}
    \begin{subfigure}{0.33\textwidth}
    \includegraphics[width=\linewidth]{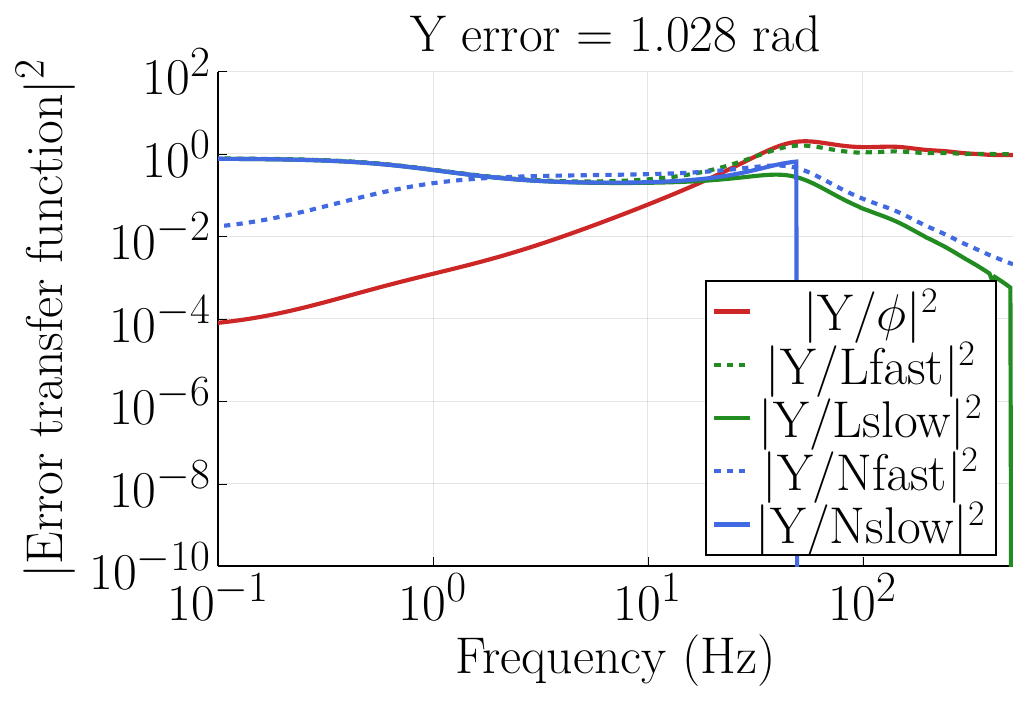}
    \caption{Error transfer functions to $Y$} \label{fig:erry_doublewfs}
    \end{subfigure}

    \medskip
    \begin{subfigure}{0.33\textwidth}
    \includegraphics[width=\linewidth]{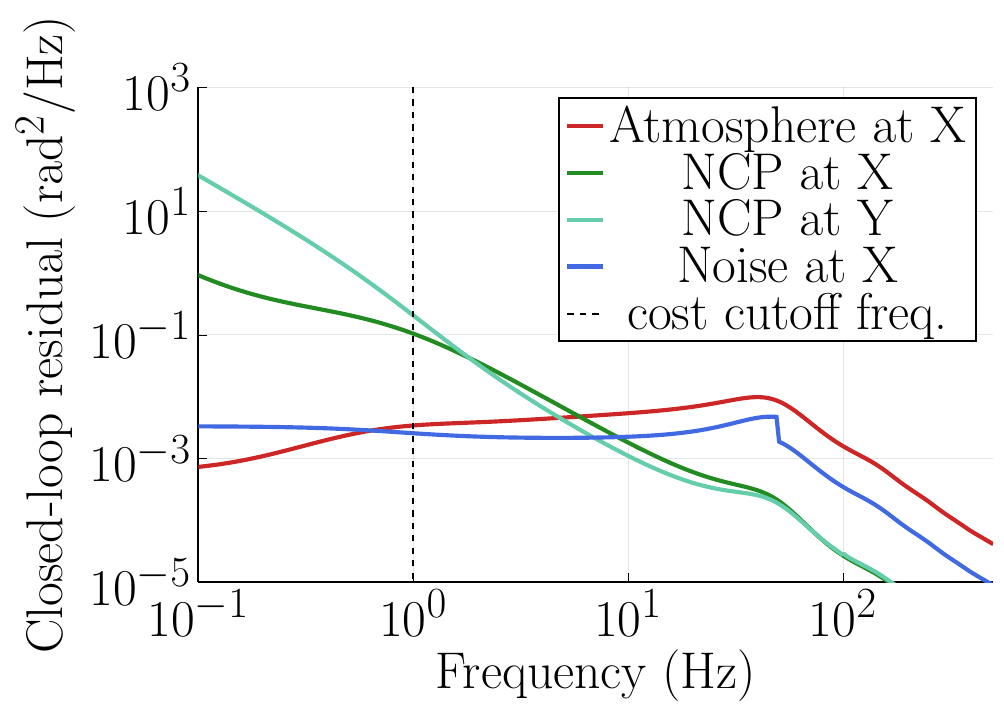}
    \caption{Closed-loop integrands} \label{fig:integrands_doublewfs}
    \end{subfigure}\hspace*{\fill}
    \begin{subfigure}{0.33\textwidth}
    \includegraphics[width=\linewidth]{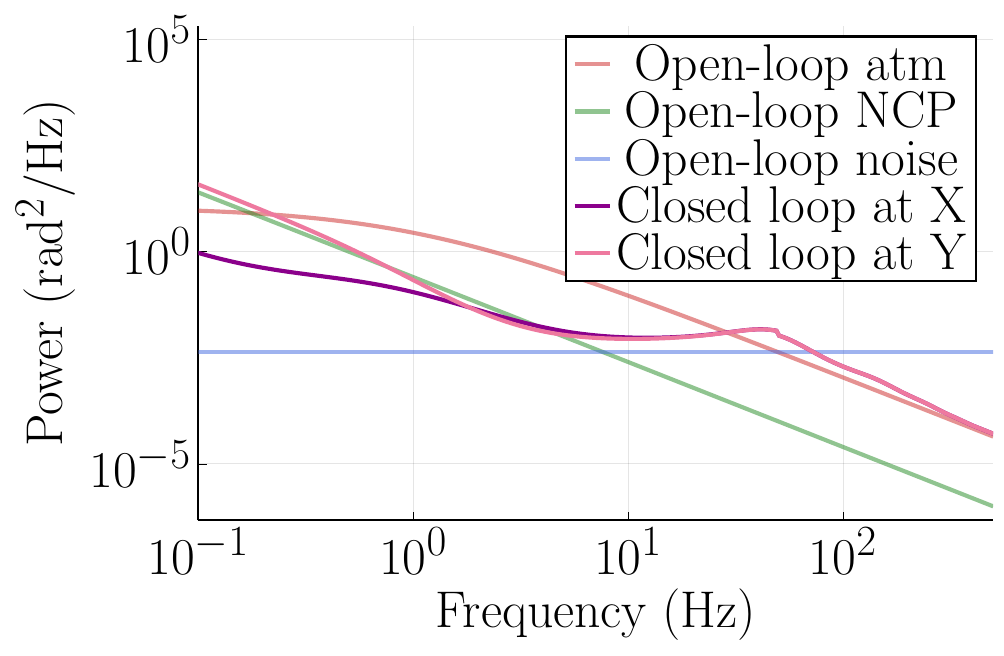}
    \caption{Open- and closed-loop PSDs} \label{fig:psds_doublewfs}
    \end{subfigure}\hspace*{\fill}
    \begin{subfigure}{0.33\textwidth}
    \includegraphics[width=\linewidth]{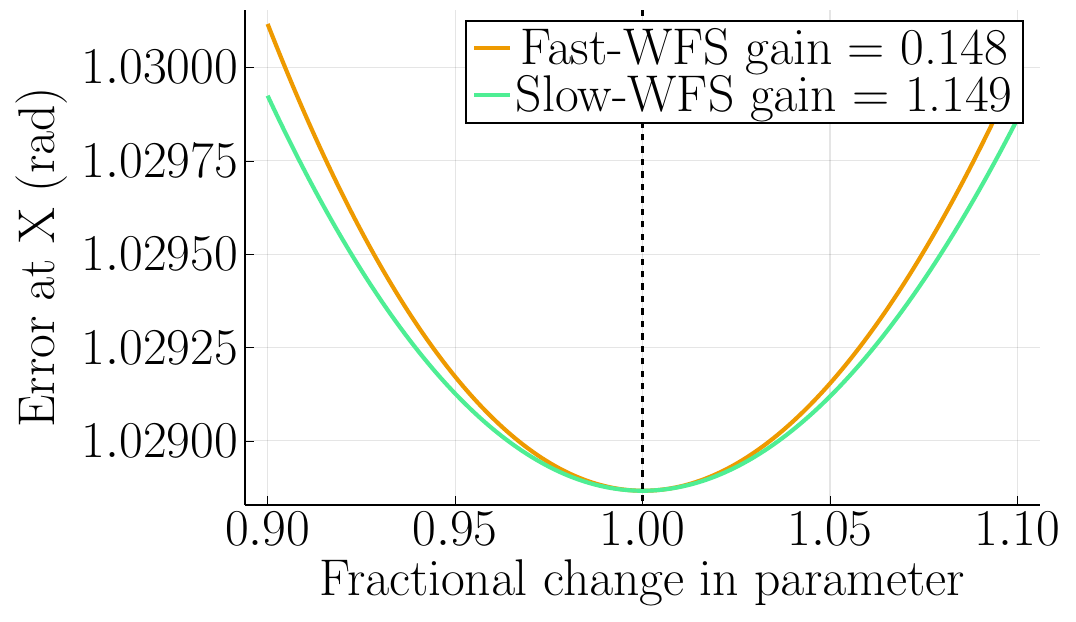}
    \caption{Error at $X$ as a function of gain} \label{fig:minplot_doublewfs}
    \end{subfigure}
    
    \caption{The double IC controller is able to correct slow NCP aberrations, but suffers from inter-arm NCP transfer. There is attenuation of slow NCP aberrations (solid green line in Figure~\ref{fig:errx_doublewfs}) but they still contribute strongly to the residual error (dark green line in Figure~\ref{fig:integrands_doublewfs}) and closed-loop performance at $X$ is still marginally worse than $Y$.} \label{fig:doublewfs}
\end{figure}

We observe improved correction at $X$ and degraded correction at $Y$. This can be attributed to successful correction of slow NCP aberrations at $X$ (Figure~\ref{fig:errx_doublewfs}), and significant transfer of slow NCP aberrations and slow-WFS noise to $Y$ (Figure~\ref{fig:erry_doublewfs}). In terms of closed-loop error contributions, the gap between NCP aberrations from $X$ and $Y$ becomes significantly smaller, but they remain the most significant component of the overall error (Figure~\ref{fig:integrands_doublewfs}). This motivates the addition of the high-pass filter to the fast controller, in order to remove the transfer of NCP aberrations between WFSs.

\subsection{Double integrator control with filtering}

As above, we optimize over the gains for both integrators, and add optimization over the cutoff frequency for the high-pass filter.

\begin{figure}
    \begin{subfigure}{0.33\textwidth}
    \includegraphics[width=\linewidth]{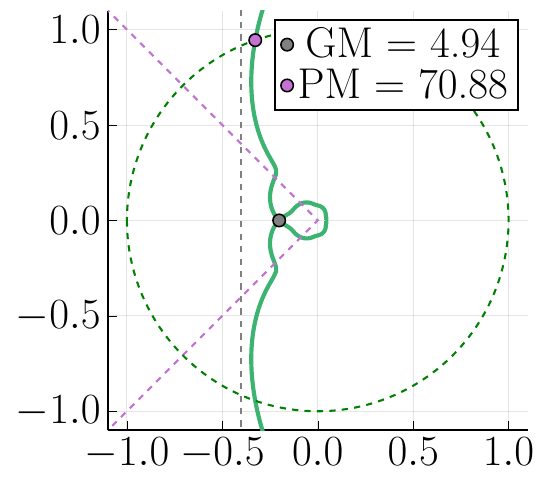}
    \caption{Nyquist diagram} \label{fig:nyquist_doublewfs_hpf}
    \end{subfigure}\hspace*{\fill}
    \begin{subfigure}{0.33\textwidth}
    \includegraphics[width=\linewidth]{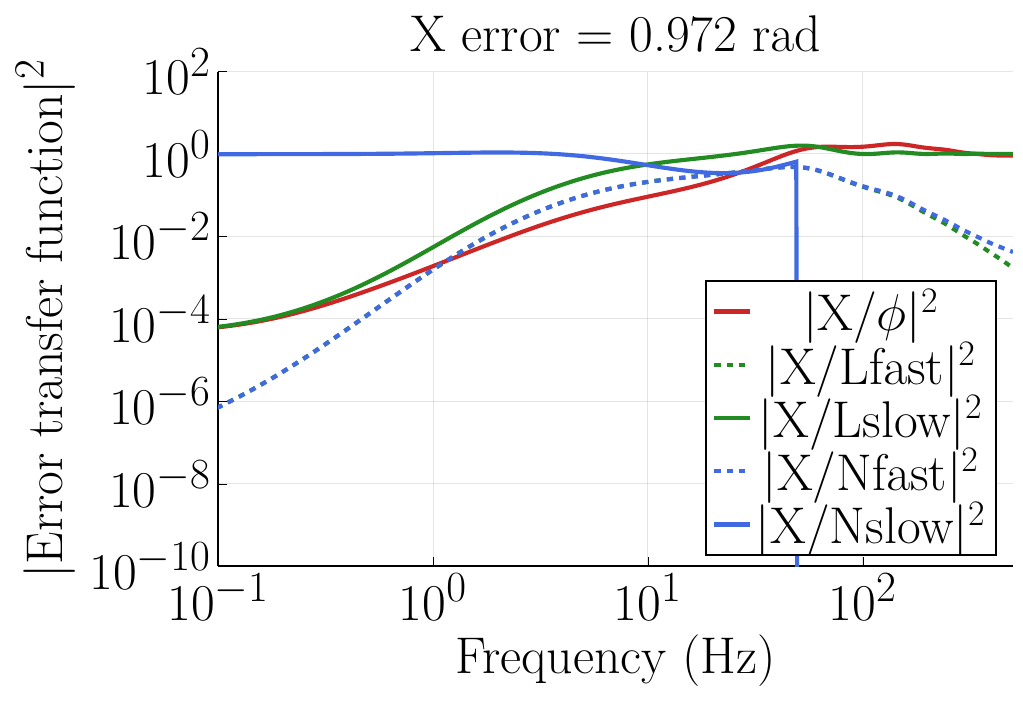}
    \caption{Error transfer functions to $X$} \label{fig:errx_doublewfs_hpf}
    \end{subfigure}\hspace*{\fill}
    \begin{subfigure}{0.33\textwidth}
    \includegraphics[width=\linewidth]{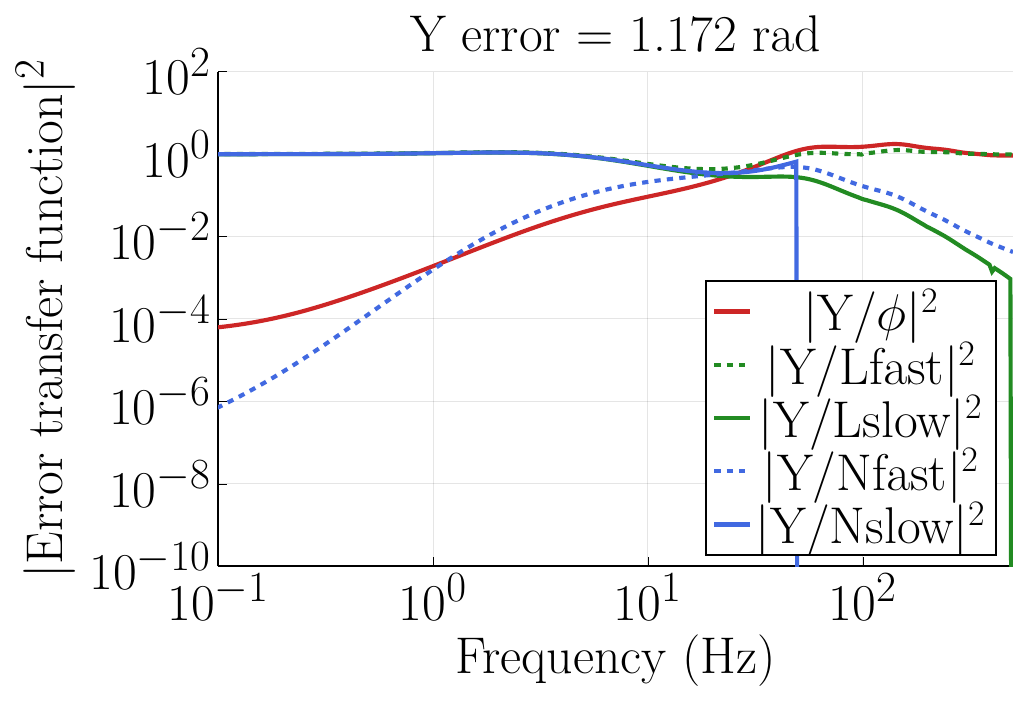}
    \caption{Error transfer functions to $Y$} \label{fig:erry_doublewfs_hpf}
    \end{subfigure}

    \medskip
    \begin{subfigure}{0.33\textwidth}
    \includegraphics[width=\linewidth]{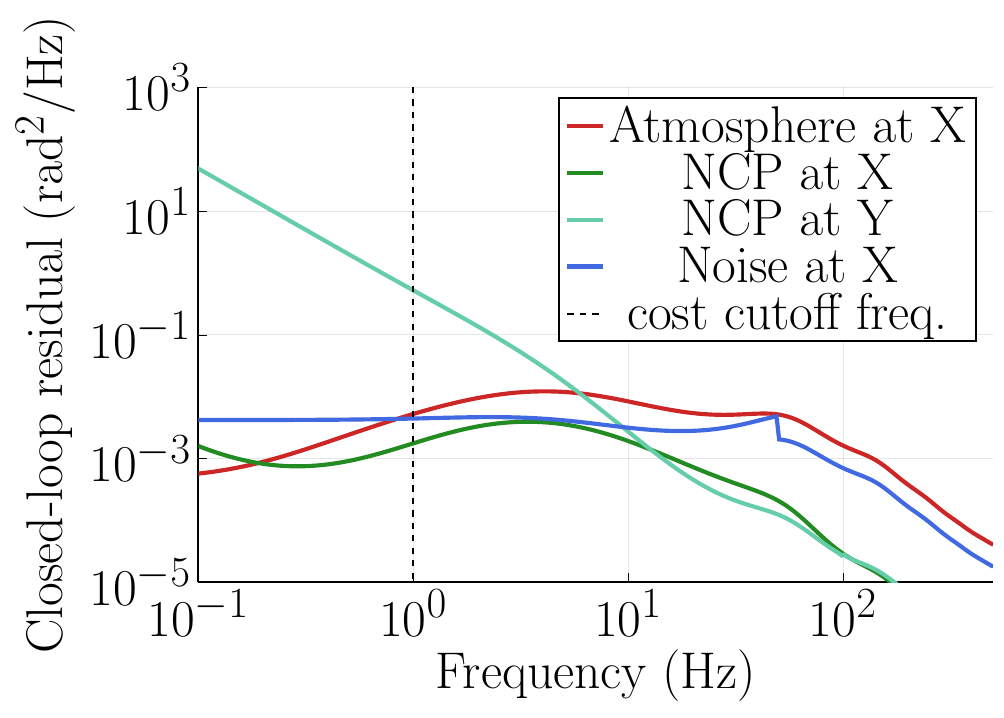}
    \caption{Closed-loop integrands} \label{fig:integrands_doublewfs_hpf}
    \end{subfigure}\hspace*{\fill}
    \begin{subfigure}{0.33\textwidth}
    \includegraphics[width=\linewidth]{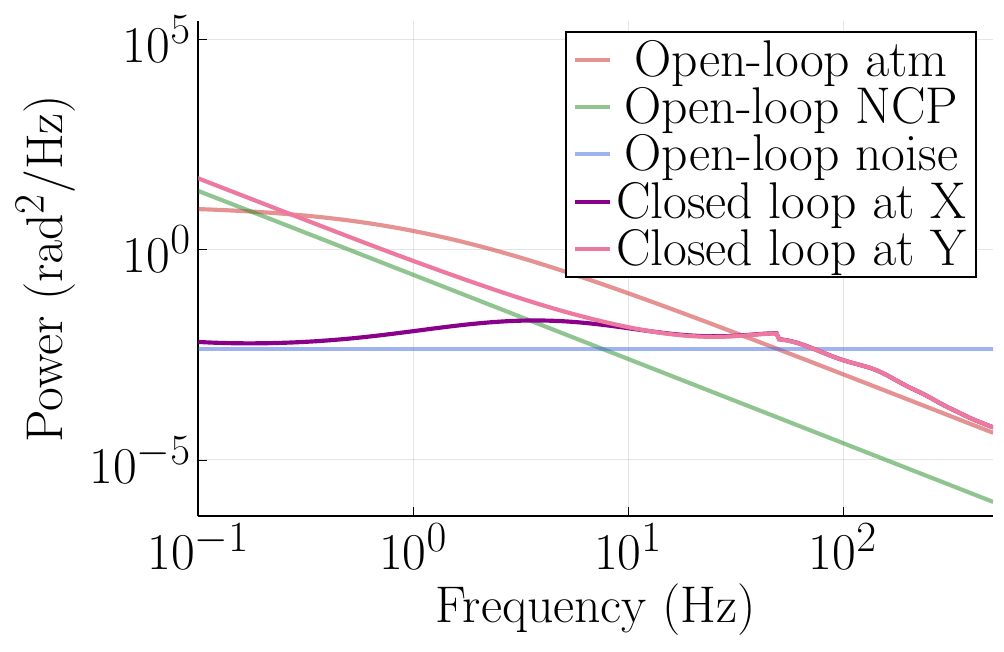}
    \caption{Open- and closed-loop PSDs} \label{fig:psds_doublewfs_hpf}
    \end{subfigure}\hspace*{\fill}
    \begin{subfigure}{0.33\textwidth}
    \includegraphics[width=\linewidth]{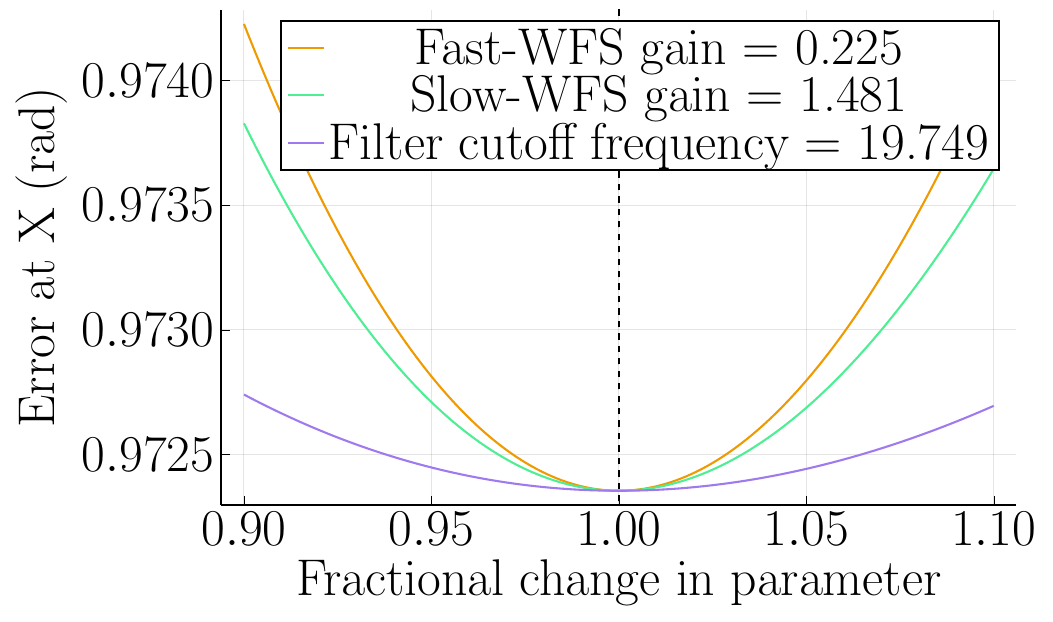}
    \caption{Error at $X$ as a function of gain} \label{fig:minplot_doublewfs_hpf}
    \end{subfigure}
    
    \caption{Double IC-HPF corrects slow NCP aberrations and attenuates inter-arm NCP transfer. Slow NCP and fast noise are attenuated significantly more (solid green and dashed blue lines in Figure~\ref{fig:errx_doublewfs_hpf}). This results in the residuals from both dropping below the atmospheric residual (dark green and blue lines relative to red line in Figure~\ref{fig:integrands_doublewfs_hpf}) and therefore in significantly better closed-loop performance at $X$ relative to $Y$ (pink and purple lines in Figure~\ref{fig:psds_doublewfs_hpf}.)} \label{fig:doublewfs_hpf}
\end{figure}

We observe improved correction at $X$ due to less transfer of NCP aberrations from $Y$. Figure~\ref{fig:integrands_doublewfs_hpf} shows a significantly lower contribution due to NCP aberrations to the total error at $X$, compared both to that at $Y$ and to the error at $X$ in the double IC case.  

It is important to note that although \cite{Gerard2023} and \cite{Sengupta2024} previously used a high pass-filter with a similarly-motivated goal of reducing inter-arm NCP transfer, the assumption in this previous work was that the filter cutoff frequency would be set at the Nyquist limit of the slower WFS, whereas here this is actually not the case for the optimally-derived cutoff frequency.

\subsection{Double LQG integrator control with filtering}

We optimize over three parameters: the process noise for the slow and fast LQG controller, and the cutoff frequency on the HPF. We set $\alpha_\text{LQG} = 0.999$ for both controllers to match the leak in the IC cases.

\begin{figure}
    \begin{subfigure}{0.33\textwidth}
    \includegraphics[width=\linewidth]{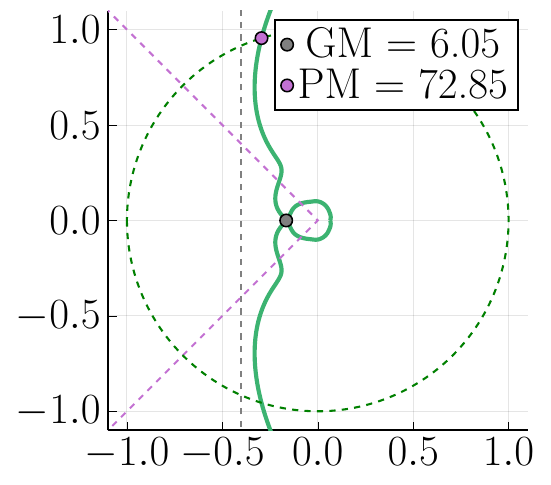}
    \caption{Nyquist diagram} \label{fig:nyquist_lqgic_doublewfs_hpf}
    \end{subfigure}\hspace*{\fill}
    \begin{subfigure}{0.33\textwidth}
    \includegraphics[width=\linewidth]{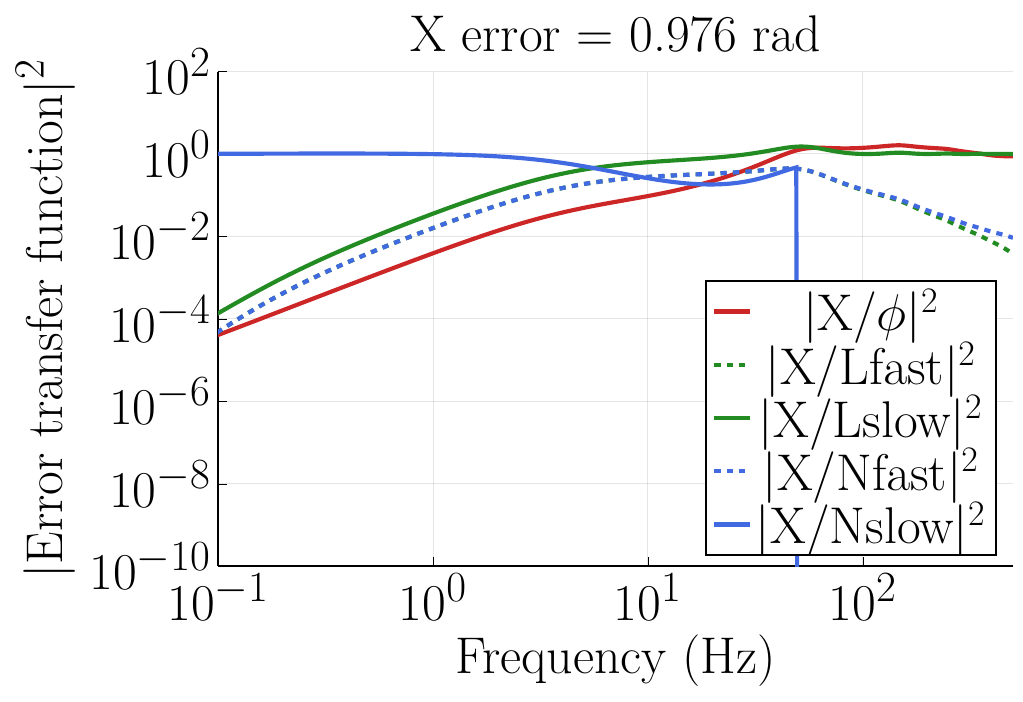}
    \caption{Error transfer functions to $X$} \label{fig:errx_lqgic_doublewfs_hpf}
    \end{subfigure}\hspace*{\fill}
    \begin{subfigure}{0.33\textwidth}
    \includegraphics[width=\linewidth]{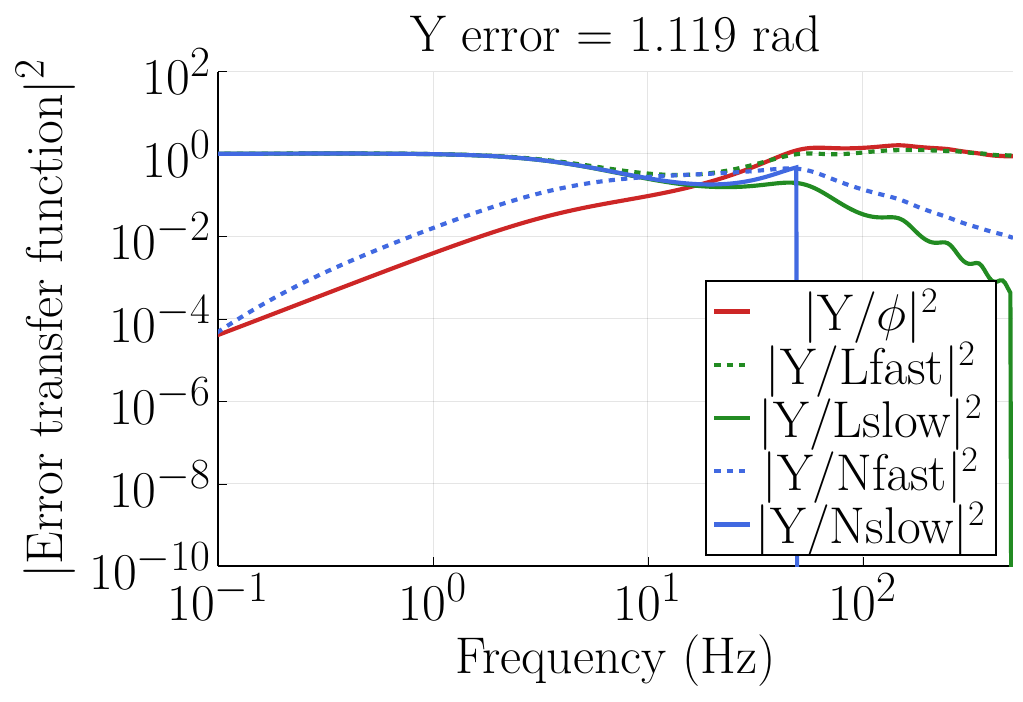}
    \caption{Error transfer functions to $Y$} \label{fig:erry_lqgic_doublewfs_hpf}
    \end{subfigure}

    \medskip
    \begin{subfigure}{0.33\textwidth}
    \includegraphics[width=\linewidth]{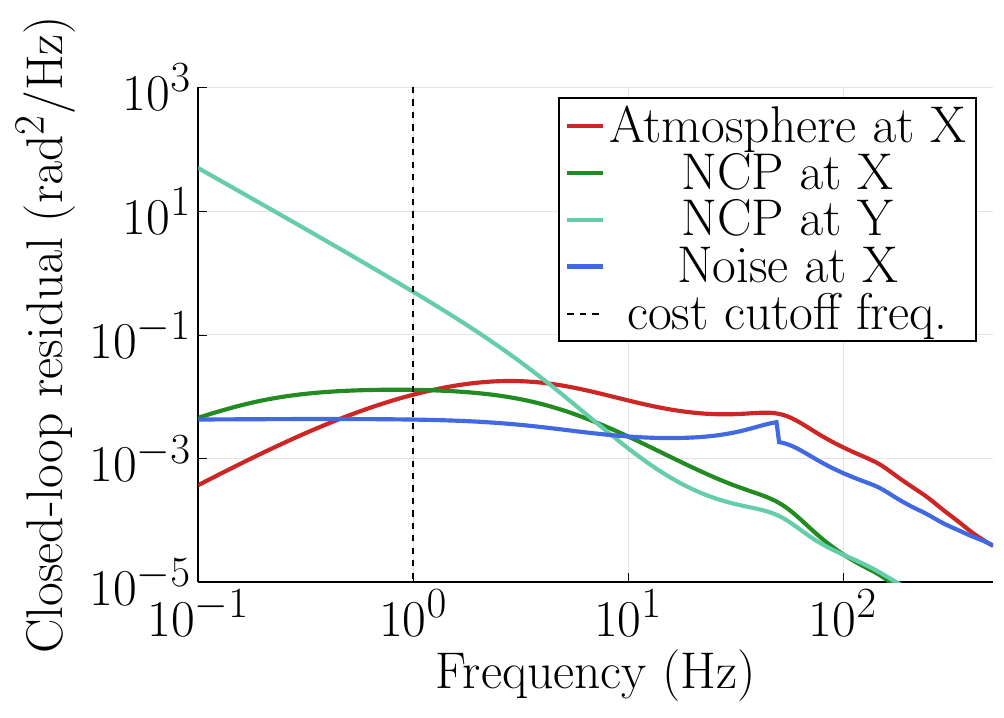}
    \caption{Closed-loop integrands} \label{fig:integrands_lqgic_doublewfs_hpf}
    \end{subfigure}\hspace*{\fill}
    \begin{subfigure}{0.33\textwidth}
    \includegraphics[width=\linewidth]{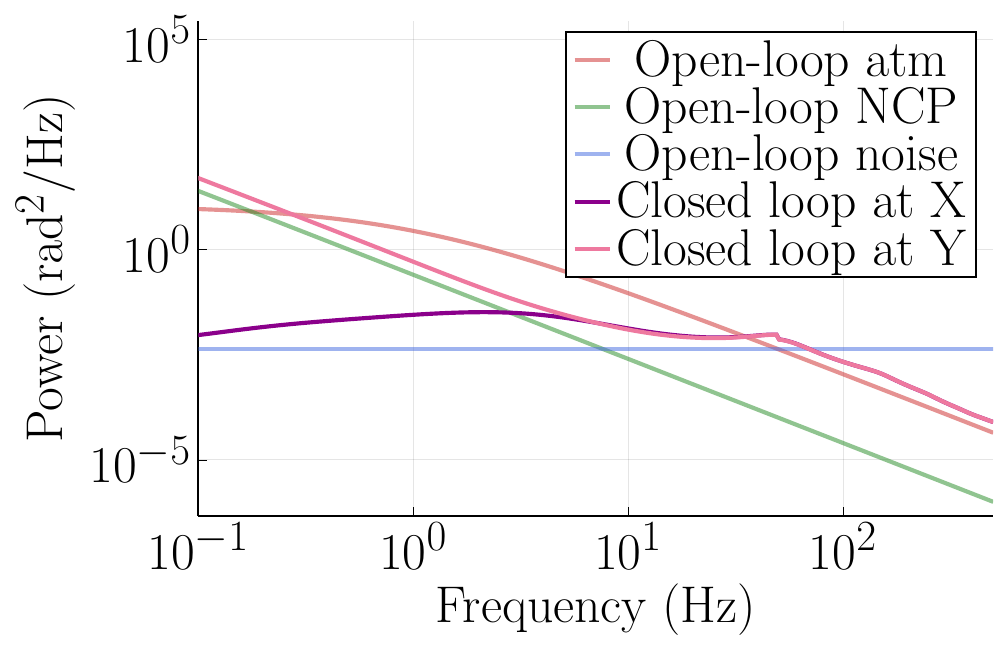}
    \caption{Open- and closed-loop PSDs} \label{fig:psds_lqgic_doublewfs_hpf}
    \end{subfigure}\hspace*{\fill}
    \begin{subfigure}{0.33\textwidth}
    \includegraphics[width=\linewidth]{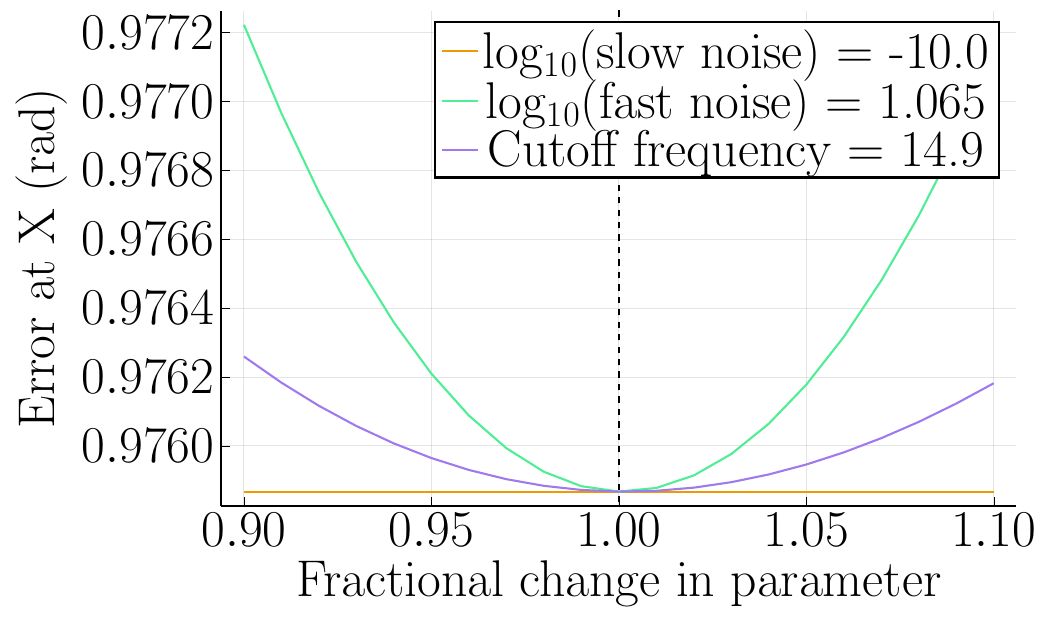}
    \caption{Error at $X$ as a function of gain} \label{fig:minplot_lqgic_doublewfs_hpf}
    \end{subfigure}
    
    \caption{Double LQG-IC-HPF performs slightly worse than double IC-HPF and does not perform as well on NCP aberrations. We observe similar trends to Figure~\ref{fig:doublewfs_hpf}, but with slightly worse rejection of slow NCP and fast noise (solid green and dashed blue lines in Figure~\ref{fig:errx_lqgic_doublewfs_hpf}). This results in a greater contribution from both to the closed-loop residual (dark green and blue lines relative to red in Figure~\ref{fig:integrands_lqgic_doublewfs_hpf}) and worse performance at $X$. The gap between the CL PSDs at $X$ and $Y$ is narrower (purple and pink lines in Figure~\ref{fig:psds_lqgic_doublewfs_hpf}).} \label{fig:lqgic_doublewfs_hpf}
\end{figure}

We observe marginal degradation in performance relative to the double IC-HPF case, with a closed-loop error at $X$ of 0.972 rad in the double IC-HPF case and 0.976 rad in the double LQG-IC-HPF case for NCP $r_0 = 1$m and $f_\text{cross}$ = 50 Hz. NCP aberrations at $X$ are a larger component of the error, and the low-frequency ETFs (in particular $X/L_\text{slow}$ and $X/N_\text{fast}$) appear to provide significantly worse correction compared to double IC-HPF. The IC attenuates disturbances to a greater extent down to lower frequencies, whereas with LQG-IC the ETFs level off more quickly at around $0.1$ Hz. This is attributable to the optimization disregarding frequencies below 1 Hz; a slight improvement in the atmospheric correction as seen by both WFSs has a more significant impact on the overall closed-loop error than the greater NCP correction being performed in the double IC-HPF case. This is consistent with the error at $Y$ also decreasing. 

We interpret this degradation as a limitation of LQG being used twice to make separate integral controllers, rather than being able to incorporate knowledge about both WFSs into one controller or to parameterize a wider family of controllers. { We therefore consider this a null result and find that initial attempts at incorporating improved controllers in the dual-WFS single-conjugate AO problem using a 1D-turbulence-based LQG controller was unsuccessful. However, better ways of exploring the space of controllers may improve performance.} More flexible LQG controllers, { consisting of more free parameters and using state-space model components that are better suited to this problem} may be able to outperform the double IC-HPF scheme. { These} controllers would have to be tailored to the dominant residual error components and would have to be optimized over a wider parameter space. We discuss this possibility further in \S\ref{sec:improving} and \S\ref{sec:further_work}.

\subsection{Interpreting parameters for the double IC-HPF case}

On-sky operation of these controllers will likely require real-time optimization of parameters as a function of changing conditions and/or observed closed-loop performance. It is therefore useful to be able to interpret which parameter controls which aspect of the closed-loop error. In the double IC-HPF case, we successfully reduce inter-arm NCP transfer, and the impact of the parameters on the individual error terms can be characterized relatively clearly. We therefore take this as our nominal case for further testing, and we further discuss the interpretation of the three parameters. 

\begin{figure}
    \centering
    \begin{subfigure}{0.32\textwidth}
        \includegraphics[width=\linewidth]{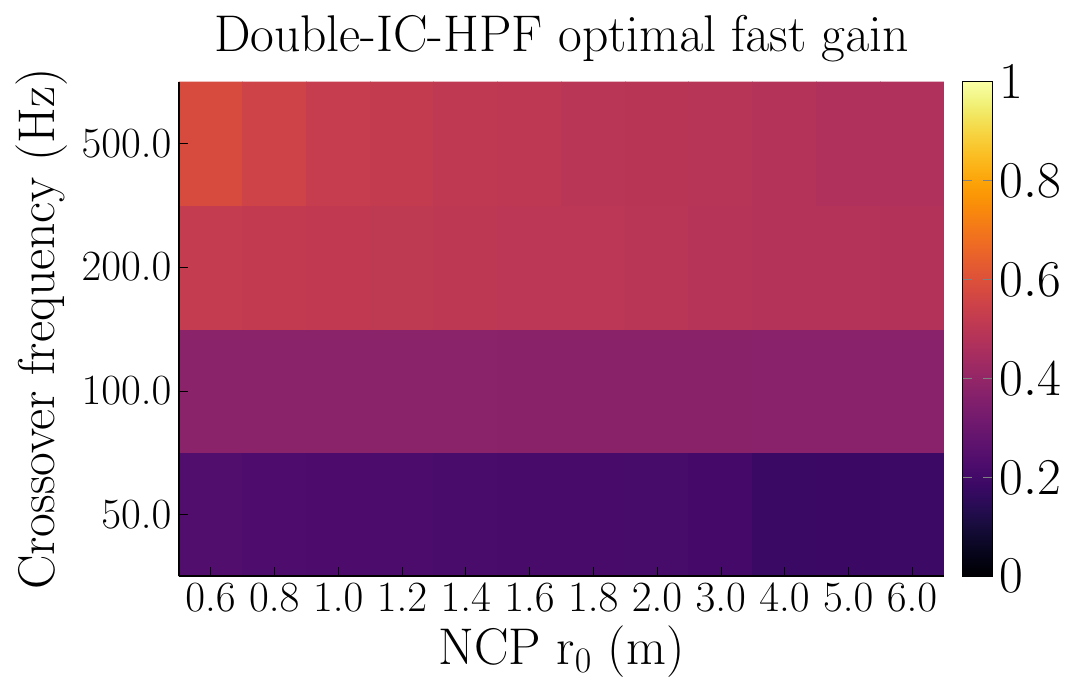}
        \caption{Atmospheric ETF} \label{fig:doublewfshpf_fastgain_heatmap}
        \end{subfigure}\hspace*{\fill}
        \begin{subfigure}{0.32\textwidth}
        \includegraphics[width=\linewidth]{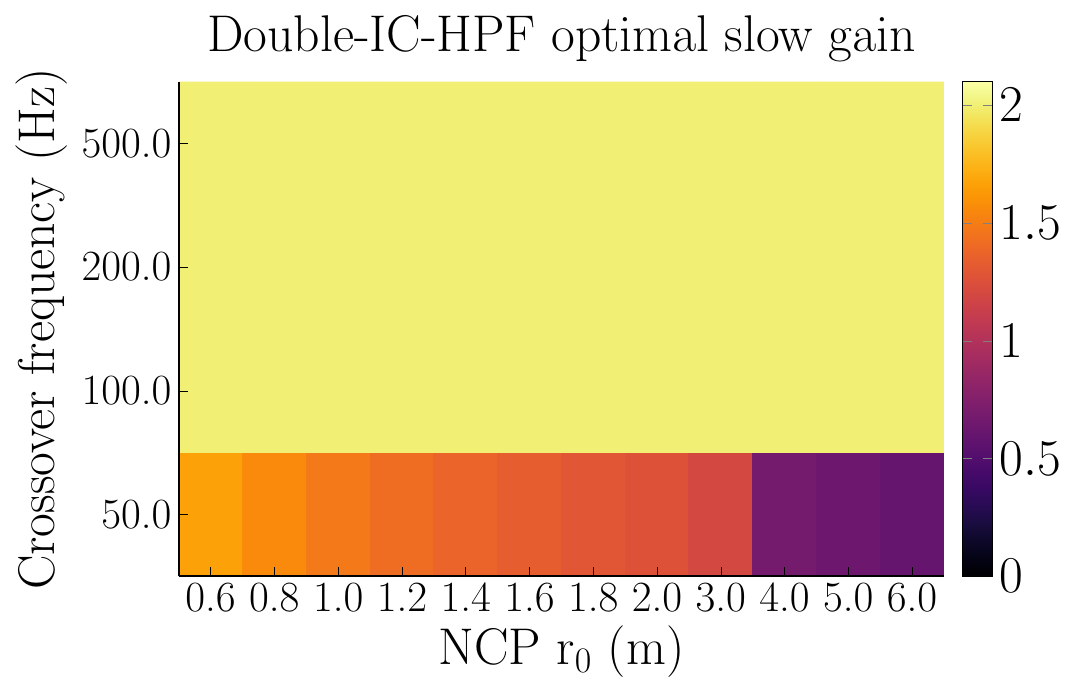}
        \caption{Error transfer functions to $X$} \label{fig:doublewfshpf_slowgain_heatmap}
        \end{subfigure}
        \begin{subfigure}{0.32\textwidth}
        \includegraphics[width=\linewidth]{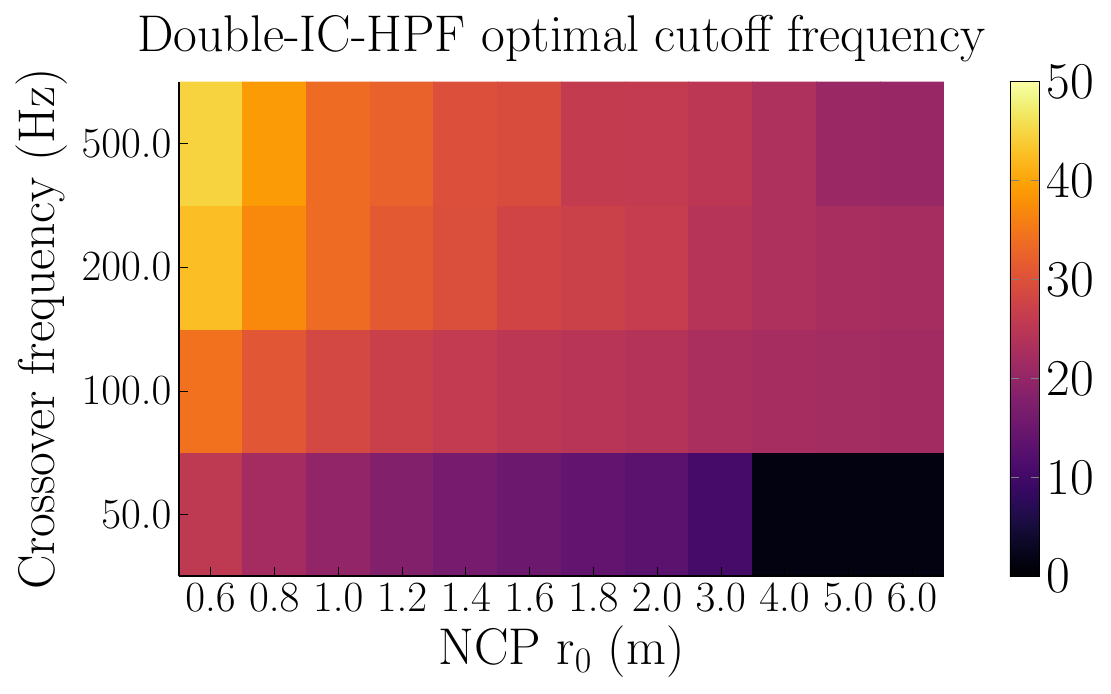}
        \caption{Error transfer functions to $X$} \label{fig:doublewfshpf_fcutoff_heatmap}
        \end{subfigure}
    \caption{The optimal double IC-HPF parameters can be interpreted in terms of the disturbance sources.}
    \label{fig:doublewfshpf_heatmaps}
\end{figure}

In the single-WFS case, the gain trades off between atmospheric correction and WFS noise; the higher the gain, the more the controller trusts any particular WFS measurement. Figure~\ref{fig:doublewfshpf_fastgain_heatmap} shows this holds for the fast-WFS gain in the double IC-HPF case as well. The fast-WFS gain is relatively independent of the strength of NCP errors, and gets smaller as the noise term gets larger (i.e., as the crossover frequency gets smaller).

The slow-WFS gain (Figure~\ref{fig:doublewfshpf_slowgain_heatmap}) is largely determined by the strength of NCP errors, since this is the main role of the slow WFS, and somewhat trades off against noise. The gain gets larger when NCP aberrations are stronger (i.e., their length scale is smaller), and in cases where the noise is sufficiently low the slow-WFS gain reaches 2, the maximum value imposed during the optimization.

The cutoff frequency (Figure~\ref{fig:doublewfshpf_fcutoff_heatmap}) increases with higher NCP aberrations and decreases with higher noise. With higher NCP aberrations, inter-arm transfer of NCP also gets larger and the filter therefore needs to play a greater role in suppressing it. With higher noise, the WFS measurements deviate to a greater extent from the true aberrations, meaning inter-arm NCP transfer becomes less significant than performing better correction on each arm. The role of the filter is also diminished since the gain for both controllers is lower, meaning the signal that would be transferred between arms is also reduced.

\subsection{Improving upon the optimal controller}\label{sec:improving}

The double LQG-IC-HPF results suggest that, rather than only using controllers that would perform better in the single-WFS case, a more targeted approach to further optimization is necessary. In several other cases, we observe that the cutoff frequency for the optimal controller is 0 Hz; that is, the optimal solution is to ignore the transfer of NCP aberrations between arms because this is outweighed by the marginal improvement in atmospheric and noise correction. Optimization beyond this point would likely not be able to further improve upon the common-path terms and would have to more explicitly handle the NCP aberration.

To further assess this, we optimized a controller (for our nominal parameters of NCP $r_0 = 1$m and $f_\text{cross} = 50$ Hz) with an LQG integrator and high-pass filter on the fast arm and a simple integrator on the slow arm. This optimization yielded a cutoff frequency of 0 Hz, so we compared this to two other cases: the optimized double IC-HPF shown in Figure~\ref{fig:doublewfs_hpf}, and a hybrid controller using the integrators from the fast-LQG-IC-HPF case and the filter from the double IC-HPF case.

\begin{figure}
    \centering
    \begin{subfigure}{0.49\textwidth}
        \includegraphics[width=\linewidth]{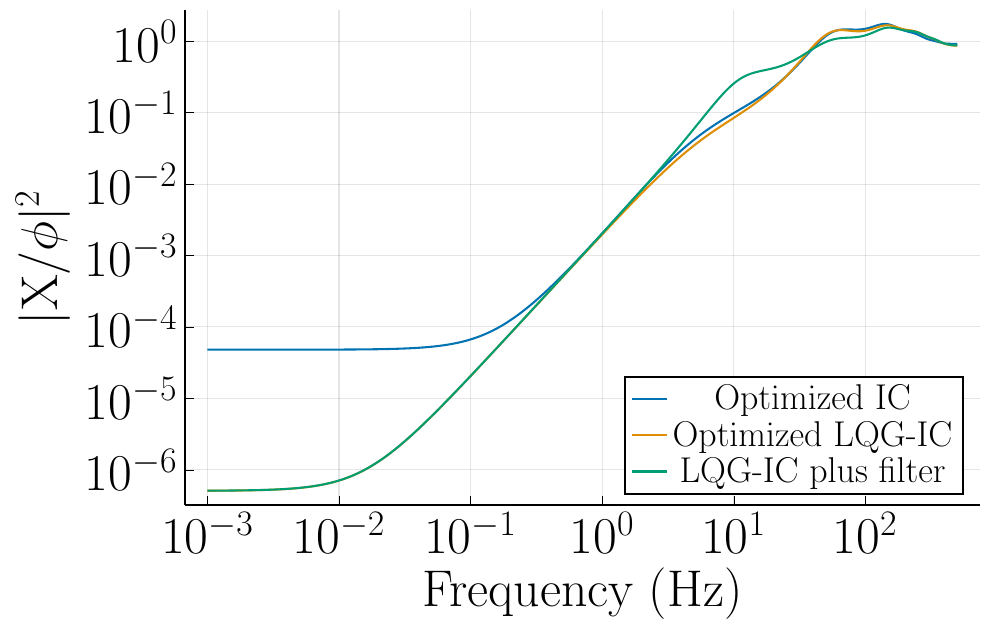}
        \caption{Atmospheric ETF} \label{fig:dropping_filter_atm}
        \end{subfigure}\hspace*{\fill}
        \begin{subfigure}{0.49\textwidth}
        \includegraphics[width=\linewidth]{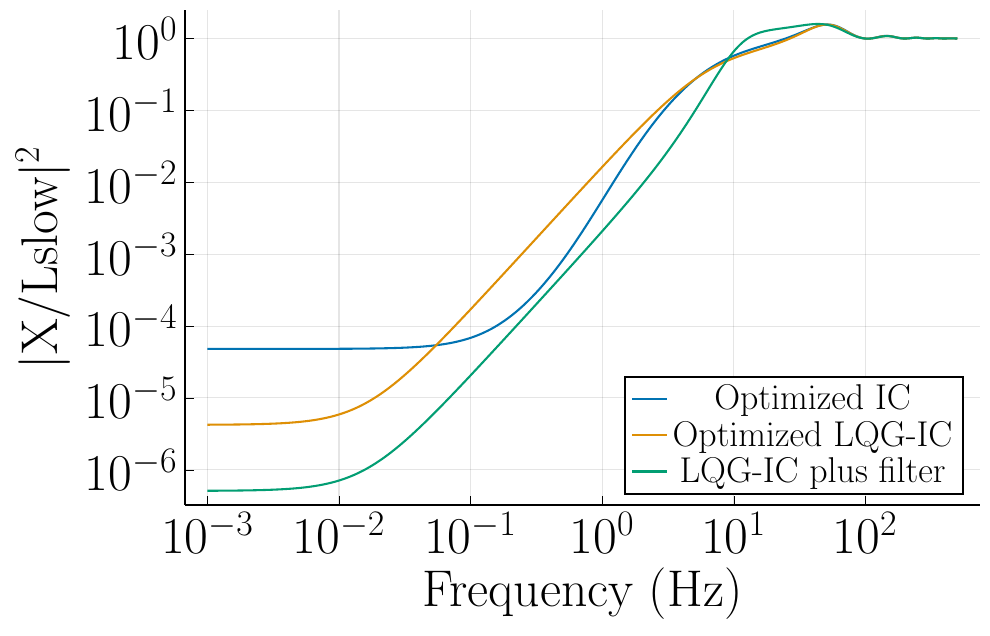}
        \caption{Slow non-common-path ETF} \label{fig:dropping_filter_ncp}
        \end{subfigure}
    \caption{Error transfer functions for three controllers in a case in which no handling of NCP transfer provides the optimal controller.}
    \label{fig:dropping_filter}
\end{figure}

Figure~\ref{fig:dropping_filter} shows the ETFs for these three cases. LQG-IC appears to improve the steady-state correction of atmospheric aberrations by around two orders of magnitude, and to slightly improve upon the IC atmospheric correction in the region past 1 Hz. Correction of slow NCP is slightly worse due to the lack of filtering. In the case where we add in the filter from the double IC-HPF, we see worse correction of $>1$ Hz atmospheric aberrations, and slightly better correction of $1-10$ Hz NCP aberrations, with a sharper overshoot past 10 Hz. However, the static NCP correction improves over both optimized cases by an order of magnitude; this is not considered within the error metric because it would require time-series runs of around 1000 seconds for its effect to be seen.


{ This result suggests that control strategies that would yield improvements in the one-WFS case, considering only atmospheric and noise components, do not necessarily perform better in the two-WFS case when also considering non-common-path errors. This is because such controllers are not necessarily able to correct the additional error induced due to inter-arm NCP transfer. Controller designs for the two-WFS one-DM problem are therefore required to specifically address the NCP aberration components as a significant part of the overall error. Having observed no improvement in control due to the LQG-IC-HPF strategies, we consider only double IC-HPF for the remainder of this work.}

Finally, we note that the separation between the error at $X$ and $Y$ remains relatively small ($\sim$10\%) in all cases. This is largely because an efficient way to reduce error seen by $X$ is to reduce the contributions from the common-path terms. However, reducing error at $Y$ is not a necessary outcome, as long as the aberration seen by the fast WFS is within its dynamic range. This presents slack in the optimization problem that could be exploited by controllers that are more finely tuned for this particular problem: a potential avenue for further optimization would be to deliberately degrade control at $Y$ in order to provide better control at $X$. This is consistent with the gap between closed-loop error at $X$ and $Y$ becoming smaller when going from double IC-HPF to double LQG-IC-HPF. Since control at $Y$ improved at the cost of more slow NCP correction, the gap became smaller and performance at $X$ became slightly worse; a different control architecture could make use of this in the other direction in order to widen the gap and provide better performance at $X$. We have not considered this further in order to work with parameters that have a clear interpretation in terms of the disturbance sources.

\section{Time-domain performance}\label{sec:time}

We characterized the performance of these controllers in the time domain using a two-dimensional simulator that generates and corrects full phase screens.

\subsection{Time-domain simulator setup}
\label{sec:2dsims}
Time domain simulations were conducted using Lawrence Livermore's
end-to-end AO simulation tool. This python tool is an updated version
of the code used in the development of the Gemini Planet Imager and
advanced control algorithms, e.g. \cite{Poyneer2007}. The simulation
models the atmosphere with a set of thin layers of Kolmogorov turbulence, which evolve
using auto-regression and frozen flow \citep{Srinath2015}. The non-common-path
phase disturbances in each wavefront sensor leg 
also have atmospheric statistics, but with a negligible wind velocity and 
much longer coherence length ($r_0$) to model air turbulence on the optical bench.

The AO simulation operates with several temporal sub-steps per AO frame to
capture both atmospheric dynamics and fractional-frame computational delays 
in the control system. For this study the main AO loop ran at 1 kHz, with five 
sub-steps per AO frame. The slow WFS was modeled as integrating over 10 
full AO frames. For simplicity and to isolate the temporal behaviors under 
exploration here, both the fast and slow WFSs were Shack-Hartmann (SH)
sensors operating at 800 nm, with fifteen subapertures across a 3-m primary mirror. 
Both SH sensors had pixels of angular size $\lambda/d$ ($d = D/15$). Fourier optics
was used to generate the integrated pixel values for each $4\times4$ pixel subimage. Photon and read noise were added to the pixels based on reasonable detector characteristics and guide star brightness photometry.
Wavefront slopes (spot locations) were determined with matched filtering, then 
the phase estimate was calculated with Fourier-Transform wavefront 
reconstruction \citep{Poyneer2005}.

The temporal performance of the AO simulator was validated using the standard 
ETF estimation technique on open-loop and closed-loop measurements, i.e. 
see Section 4.D of \cite{Poyneer2016}. The AO simulator performance was
compared to the analytic models given above for the transfer functions, verifying
that the parameter methods that rely on these same models are appropriate
for use in determining the controllers to use in the AO simulation study.

Additionally, the open-loop measurement PSDs, 
averaged over each actuator, were used to determine the crossover point between the
atmospheric turbulence input and the WFS noise (primarily due to photon noise). 
A stellar V magnitude of 5.5 produced a crossover frequency of 200 Hz for 
atmospheric turbulence with $r_0 = 10.3$ cm.

Performance is evaluated in terms of in-band error at 800 nm as seen by in the slow WFS leg. This is calculated from the high-resolution phase in the simulation (not the slow WFS measurements), which is low-pass filtered to remove uncontrollable spatial frequencies (e.g. fitting error). The in-band RMS is calculated as the spatial standard deviation of the filtered phase, so it includes temporal and noise error and also aliasing. These additional sources of error mean we expect the closed-loop error from this simulation to always be higher than the corresponding setting in the analytic frequency-domain model. We simulate closed-loop control at each setting 10 times for 0.1s each and report the mean value as well as the first and third quartiles. 

\subsection{Time-domain results}

\begin{figure}
    \centering
    \includegraphics[width=\textwidth]{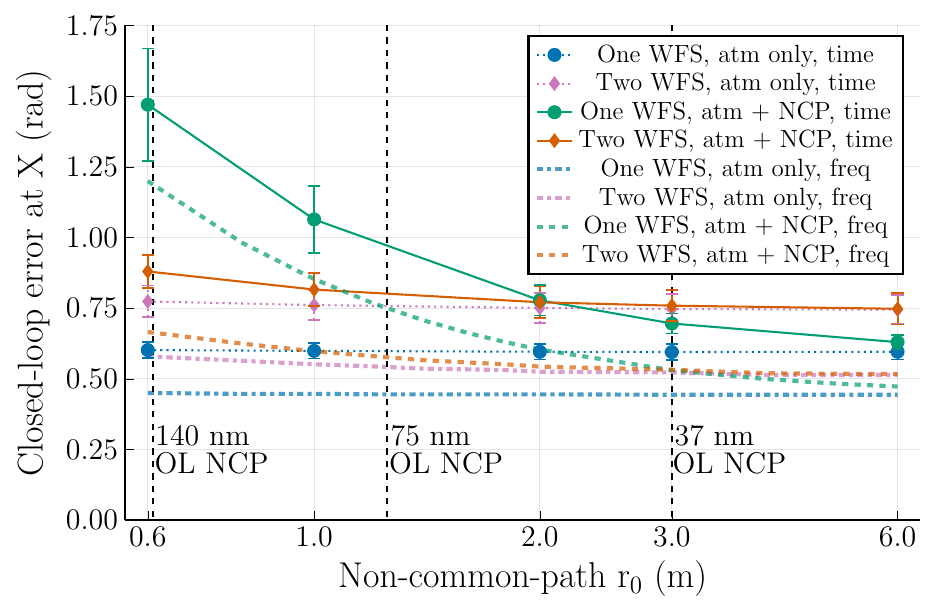}
    \caption{Two-WFS control improves performance at $X$ in the presence of strong NCP errors. The optimal closed-loop error (in rad rms) for the double IC-HPF scheme (in orange) is lower than that for the single IC (in green) when NCP aberrations are stronger (lower NCP $r_0$). Results are shown with and without NCP error, and using the frequency- and time-domain models. All simulations are done using parameters optimized at each point using the frequency-domain model.}
    \label{fig:hairdryer}
\end{figure}

Figure~\ref{fig:hairdryer} shows the closed-loop error at $X$ under the optimal single IC and the optimal double IC-HPF, as a function of NCP $r_0$ at $f_\text{cross} = 200$ Hz. The frequency-domain simulations use the atmospheric open-loop PSD estimated from the time-domain simulations. Note that since the optimal parameters are computed assuming NCP error is present, the error in the atmosphere-only cases can vary as a function of NCP $r_0$. 

In the absence of NCP error, the single integrator always outperforms the dual-WFS case. This is expected, since the double IC-HPF degrades atmospheric correction in order to provide NCP correction that is not necessary in this case. When NCP error is included, the single integrator is significantly worse. The double IC-HPF provides better overall correction for stronger NCP aberrations, and is only outperformed by the single integrator for NCP $r_0$ above 3m. This crossover (between the green and orange lines) represents the point at which the NCP aberrations are sufficiently weak such that correcting them has less of an impact than the better atmospheric rejection of the single integrator. In the time-domain simulation this crossover is observed at NCP $r_0$ = 2m. This difference results from the duration of the time-domain runs being 0.1s, meaning control in the 1-10 Hz range is not reflected in the time-domain points. The time-domain crossover moves to 3m when considering 1s runs. This is not shown due to the computational expense of repeating the longer runs 10 times each in order to show error bars. This indicates that there may be small outstanding discrepancies when optimizing two-WFS control in practical on-sky settings. It may be necessary to reconcile these in ways that are specific to each experiment.

This demonstrates that our frequency-domain simulation and optimization is representative of the more complete time-domain simulation, and that the moving-average model and the reduction from full phase screens to single variables for each disturbance were not significant enough changes to greatly alter the overall outcome. We therefore find that in the presence of non-common-path aberrations, two-WFS control can result in lower wavefront error at the science imaging plane, and that control can be optimized with respect to empirically estimated power spectra using a computationally inexpensive analytic model.

\section{Options for further optimal control research}\label{sec:further_work}
In this paper we focused on a model with parameters that are interpretable and that could work with minimal existing knowledge of the system. This is beneficial for technology demonstration projects where a proof of concept should not require detailed information about the AO system and the aberration sources, and where the ability to rapidly explore the space of possible controllers based on real-time observed performance is valuable. Better-performing control algorithms are achievable, but require more precise information about conditions such as the wind speed and atmospheric $C_n^2$ profile. We discuss two options for this kind of optimal control: `LQG-first' and model predictive control.

\subsection{`LQG-first' control design}
{ Achieving the accurate system identification required for LQG control to be the optimal solution} is difficult in practice, but a controller operating on a different system than the one that was identified may still yield improved performance. In the `LQG-first' approach, we do not attempt to directly model the system. Instead, we consider a flexible parameterization of a state-space model and treat its parameters as free variables that can be used to shape the resulting LQG controller, without interpreting these parameters in the context of the actual control problem. 

LQG-first allows us to explore the broader space of possible controllers, but it requires optimization over many more nonlinearly-coupled parameters at once. Demonstrating the local or global optimality of an LQG-first controller is computationally difficult. Further, it requires a broadly-defined set of potential LQG features; there is no guarantee that any particular model parameterization admits the optimal controller, so it becomes necessary to include a range of aberration components and to zero out the components that do not prove to be useful. This increases the search space drastically. It becomes difficult to identify controllers that perform well in the absence of interpretability for the free parameters. We can assess the controller error budget for any particular LQG-first controller, but its performance cannot be clearly attributed to any particular component of the assumed state-space model in the way that is enabled by the IC-HPF architecture. This lack of interpretability makes it difficult to adjust a controller to changing conditions in real time. A highly flexible state-space model together with an improved optimization procedure could allow LQG-first to shape controllers that yield better performance, but the steps to achieving this remain unclear.

\subsection{Model predictive control}
Model predictive control (MPC) allows the quadratic control objective from LQG to be augmented with linear constraints, to form a convex optimization problem to be solved at each timestep. This would yield the optimal controller for minimizing closed-loop error at $X$ while explicitly considering the dynamic-range limit at $Y$, enabling the controller to make use of the additional slack between the two points. This is suitable for real-time operation due to the existence and uniqueness guarantees for solutions to convex optimization problems, and due to the availability of high-performance open-source modeling tools for these problems \citep{cvxpy}. However, as with LQG, MPC depends strongly on accurate system identification. 

As a nonlinear method, MPC does not immediately lend itself to the frequency-domain analysis that we employ. This is not a significant drawback in the presence of good system identification, since we are not required to optimize MPC in the frequency domain. For the purposes of comparing ideal frequency-domain performance and checking stability, we can make use of explicit MPC, which states that under reasonable conditions, MPC can be expressed as a piecewise affine combination of linear controllers such as those produced by LQG \citep{Alessio2009}. 

\section{Conclusion}\label{sec:conclusion}

We { developed a frequency-domain simulation of the single-conjugate dual-WFS problem, demonstrated that it produced similar results to a high-fidelity time-domain AO control simulation}, and computed optimal parameters for { several control schemes over} a range of NCP and noise conditions. We demonstrated that filtering the fast WFS prevents inter-arm NCP transfer and reduces closed-loop error relative to the one-WFS case and the unfiltered two-WFS case. { This approach yields significantly better control at the science plane: for instance, when the non-common-path $r_0$ is 0.6m, the optimal one-WFS vs. two-WFS controller provides a closed-loop error of around 1.5 rad rms vs. 0.6 rad rms, respectively.}

We assessed control algorithms that show better performance in the single-WFS case and observed inconsistent results with respect to the the dual-WFS metric. { For instance, when designing a controller using an LQG integrator on only one arm, we found the optimization favored no temporal filtering (\S\ref{sec:improving}). The optimal parameters for this control configuration required that we neglect inter-arm NCP transfer, suggesting that controllers that are generally known to provide improved performance in the single-WFS case do not necessarily provide the same improvement for the dual-WFS case. In particular, we find that integrator control with a high-pass filter on the fast WFS output provides significant improvement over the single-WFS integrator, but a first attempt at LQG control does not show further improvements beyond this.} We noted the potential for control algorithms that improve performance as seen by the slow WFS by deliberately degrading performance at the fast WFS. We discussed two options for further performance improvement { based on more specific state-space modeling in order to rigorously impose constraints on the total wavefront error seen by a WFS with dynamic range limits.}

With the control algorithms presented in this work, focal-plane WFS can be implemented at existing observatories with no hardware changes beyond installing the WFS itself.

The code and data underlying the results presented in this paper are publicly available at \href{https://github.com/aditya-sengupta/multiwfs}{https://github.com/aditya-sengupta/multiwfs}.

\begin{acknowledgments}
AS thanks Vincent Chambouleyron, Emiel Por, and Parth Nobel for consultations on implementing the control algorithms and metrics. We acknowledge funding from the Lawrence Livermore National Laboratory LDRD Program (24-LW-002 and 25-ERD-003); U.S. Department of Energy; National Nuclear Security Administration (DE-AC52-07NA27344). This document was prepared as an account of work sponsored by an agency of the United States government. Neither the United States government nor Lawrence Livermore National Security, LLC, nor any of their employees makes any warranty, expressed or implied, or assumes any legal liability or responsibility for the accuracy, completeness, or usefulness of any information, apparatus, product, or process disclosed, or represents that its use would not infringe privately owned rights. Reference herein to any specific commercial product, process, or service by trade name, trade- mark, manufacturer, or otherwise does not necessarily constitute or imply its endorsement, recommendation, or favoring by the United States government or Lawrence Livermore National Security, LLC. The views and opinions of authors expressed herein do not necessarily state or reflect those of the United States government or Lawrence Livermore National Security, LLC, and shall not be used for advertising or product endorsement purposes. The document number is LLNL-JRNL-2005648-DRAFT and the code number is LLNL-CODE-2000606.
\end{acknowledgments}

\begin{contribution}

AS created the frequency-domain simulation with input from all authors, computed the performance metrics and optimal parameters for each controller family, and led manuscript writing other than section 4.1. LP carried out the time-domain simulations and wrote section 4.1. BG conceptualized the project. All authors contributed to writing the manuscript.

\end{contribution}

\software{ControlSystems.jl \citep{ControlSystems}, DSP.jl \citep{dspjl}, PGFPlotsX.jl \citep{pgfplotsxjl}}

\appendix
\section{Transfer function derivations}

    We derive transfer functions between each input ($\phi$, $L_{\text{fast}}$, $L_{\text{slow}}$, $N_{\text{fast}}$, $N_{\text{slow}}$) and output ($X$, the signal seen by the slow WFS; and $Y$, the signal seen by the fast WFS). We do this by writing down relationships between the intermediate named signals, and eliminating any signals other than the input and output.

    Going ``backwards'', we have

    \begin{align}
        \epsilon &= \phi - D \\
        D &= \parens{\frac{1 - e^{-sT}}{sT}} e^{-sT} \parens{C_{\text{fast}} r_{\text{fast}} + C_{\text{slow}} r_{\text{slow}}} \\
        r_{\text{fast}} &= N_{\text{fast}} + \frac{1 - e^{-sT}}{sT} Y \\
        Y &= L_{\text{fast}} + \epsilon \\
        r_{\text{slow}} &= N_{\text{slow}} + \frac{1 - e^{-sRT}}{sRT} X \\
        X &= L_{\text{slow}} + \epsilon \\
    \end{align}

    To calculate $X/\phi$ and equivalently $Y/\phi$, we set the $L$ and $n$ terms to zero. We're left with $X = Y = \epsilon$.

    \begin{align}
        X &= \phi - D \\
        D &= \parens{\frac{1 - e^{-sT}}{sT}} e^{-sT} \parens{C_{\text{fast}} r_{\text{fast}} + C_{\text{slow}} r_{\text{slow}}} \\
        r_{\text{fast}} &= \frac{1 - e^{-sT}}{sT} X \\
        r_{\text{slow}} &= \frac{1 - e^{-sRT}}{sRT} X \\
    \end{align}

    This simplifies to

    \begin{align}
        X &= \phi - \parens{\frac{1 - e^{-sT}}{sT}} e^{-sT} \parens{C_{\text{fast}} \frac{1 - e^{-sT}}{sT} + C_{\text{slow}} \frac{1 - e^{-sRT}}{sRT}} X\\
        X &\parens{1 + \parens{\frac{1 - e^{-sT}}{sT}} e^{-sT} \parens{C_{\text{fast}} \frac{1 - e^{-sT}}{sT} + C_{\text{slow}} \frac{1 - e^{-sRT}}{sRT}} } = \phi\\
        \frac{X}{\phi} &= \frac{1}{1 + \parens{\frac{1 - e^{-sT}}{sT}} e^{-sT} \parens{C_{\text{fast}} \frac{1 - e^{-sT}}{sT} + C_{\text{slow}} \frac{1 - e^{-sRT}}{sRT}}}\\
    \end{align}

    For convenience we'll use the name $\text{plant} = \parens{\frac{1 - e^{-sT}}{sT}} e^{-sT} \parens{C_{\text{fast}} \frac{1 - e^{-sT}}{sT} + C_{\text{slow}} \frac{1 - e^{-sRT}}{sRT}}$. This means $X/\phi = Y/\phi = 1 / (1 + \text{plant})$.

    To calculate $X/L_{\text{fast}}$ and $Y/L_{\text{fast}}$, we set $\phi$, $L_{\text{slow}}$, and the $n$ terms to zero.

    \begin{align}
        X &= -\parens{\frac{1 - e^{-sT}}{sT}} e^{-sT} \parens{C_{\text{fast}} \frac{1 - e^{-sT}}{sT} Y + C_{\text{slow}} \frac{1 - e^{-sRT}}{sRT} X} \\
        Y &= L_{\text{fast}} + X \\
    \end{align}

    We'll first eliminate $Y$ from this:

    \begin{align}
        X &= -\parens{\frac{1 - e^{-sT}}{sT}} e^{-sT} \parens{C_{\text{fast}} \frac{1 - e^{-sT}}{sT} \parens{L_{\text{fast}} + X} + C_{\text{slow}} \frac{1 - e^{-sRT}}{sRT} X} \\
        X & \parens{1 + \parens{\frac{1 - e^{-sT}}{sT}} e^{-sT} \parens{C_{\text{fast}} \frac{1 - e^{-sT}}{sT} + C_{\text{slow}} \frac{1 - e^{-sRT}}{sRT}}} = -L_{\text{fast}} \parens{\frac{1 - e^{-sT}}{sT}}^2 e^{-sT} C_{\text{fast}}\\
        \frac{X}{L_{\text{fast}}} &= \frac{-\parens{\frac{1 - e^{-sT}}{sT}}^2 e^{-sT} C_{\text{fast}}}{1 + \text{plant}}.
    \end{align}

    Then, we'll eliminate $X$:

    \begin{align}
        Y - L_{\text{fast}} &= -\parens{\frac{1 - e^{-sT}}{sT}} e^{-sT} \parens{C_{\text{fast}} \frac{1 - e^{-sT}}{sT} Y + C_{\text{slow}} \frac{1 - e^{-sRT}}{sRT} (Y - L_{\text{fast}})} \\
        Y &\parens{1 + \text{plant}} = L_{\text{fast}} \parens{1 + \parens{\frac{1 - e^{-sT}}{sT}} e^{-sT} C_{\text{slow}} \frac{1 - e^{-sRT}}{sRT}}\\
        \frac{Y}{L_{\text{fast}}} &= \frac{1 + \parens{\frac{1 - e^{-sT}}{sT}} e^{-sT} C_{\text{slow}} \frac{1 - e^{-sRT}}{sRT}}{1 + \text{plant}}
    \end{align}

    To calculate $X/L_{\text{slow}}$ and $Y/L_{\text{slow}}$, we set $\phi$, $L_{\text{fast}}$, and the $n$ terms to zero.

    \begin{align}
        Y &= -\parens{\frac{1 - e^{-sT}}{sT}} e^{-sT} \parens{C_{\text{fast}} \frac{1 - e^{-sT}}{sT} Y + C_{\text{slow}} \frac{1 - e^{-sRT}}{sRT} X} \\
        X &= L_{\text{slow}} + Y \\
    \end{align}

    We'll first eliminate $Y$ from this:

    \begin{align}
        X - L_{\text{slow}} &= -\parens{\frac{1 - e^{-sT}}{sT}} e^{-sT} \parens{C_{\text{fast}} \frac{1 - e^{-sT}}{sT} (X - L_{\text{slow}}) + C_{\text{slow}} \frac{1 - e^{-sRT}}{sRT} X} \\
        X &\parens{1 + \text{plant}} = L_{\text{slow}} \parens{1 + \parens{\frac{1 - e^{-sT}}{sT}}^2 e^{-sT} C_{\text{fast}}}\\
        \frac{X}{L_{\text{slow}}} &= \frac{1 + \parens{\frac{1 - e^{-sT}}{sT}}^2 e^{-sT} C_{\text{fast}}}{1 + \text{plant}}.
    \end{align}

    Then, we'll eliminate $X$:

    \begin{align}
        Y &= -\parens{\frac{1 - e^{-sT}}{sT}} e^{-sT} \parens{C_{\text{fast}} \frac{1 - e^{-sT}}{sT} Y + C_{\text{slow}} \frac{1 - e^{-sRT}}{sRT} (L_{\text{slow}} + Y)} \\
        Y &\parens{1 + \text{plant}} = L_{\text{slow}} \parens{-\parens{\frac{1 - e^{-sT}}{sT}} e^{-sT} C_{\text{slow}} \frac{1 - e^{-sRT}}{sRT}}\\
        \frac{Y}{L_{\text{slow}}} &= \frac{-\parens{\frac{1 - e^{-sT}}{sT}} e^{-sT} C_{\text{slow}} \frac{1 - e^{-sRT}}{sRT}}{1 + \text{plant}}
    \end{align}

    For the $n$ transfer functions, we set $\phi, L_\text{fast}, L_\text{slow}$ to 0. This gives us $\epsilon = -D = X = Y$, so we can calculate just $X / N_\text{fast/slow}$, and the $Y$ ones will be the same.

    \begin{align}
        -X &= \parens{\frac{1 - e^{-sT}}{sT}} e^{-sT} \parens{C_{\text{fast}} r_{\text{fast}} + C_{\text{slow}} r_{\text{slow}}} \\
        r_{\text{fast}} &= N_{\text{fast}} + \frac{1 - e^{-sT}}{sT} X \\
        r_{\text{slow}} &= N_{\text{slow}} + \frac{1 - e^{-sRT}}{sRT} X \\
    \end{align}

    This gives us

    \begin{align}
        -X &= \parens{\frac{1 - e^{-sT}}{sT}} e^{-sT} \parens{C_{\text{fast}} \parens{N_{\text{fast}} + \frac{1 - e^{-sT}}{sT} X} + C_{\text{slow}} \parens{N_{\text{slow}} + \frac{1 - e^{-sRT}}{sRT} X}} \\
    \end{align}

    which simplifies to

    \begin{align}
        \begin{split}
            -X \Bigg(1 + C_\text{fast} \parens{\frac{1 - e^{-sT}}{sT}}^2 e^{-sT} &+ C_\text{slow} \parens{\frac{1 - e^{-sT}}{sT}} \parens{\frac{1 - e^{-sRT}}{sRT}}  e^{-sT}\Bigg) \\ &= \frac{1 - e^{-sT}}{sT} e^{-sT} \parens{C_\text{fast} N_\text{fast} + C_\text{slow} N_\text{slow}}
        \end{split}
    \end{align}

    This gives us the following solution for all four noise transfer functions, which is the same for fast/slow other than the controller transfer function in the numerator.

    \begin{align}
        \frac{X}{N_\text{fast/slow}} = \frac{-\frac{1 - e^{-sT}}{sT} e^{-sT} C_{\text{fast/slow}}}{1 + C_\text{fast} \parens{\frac{1 - e^{-sT}}{sT}}^2 e^{-sT} + C_\text{slow} \parens{\frac{1 - e^{-sT}}{sT}} \parens{\frac{1 - e^{-sRT}}{sRT}}  e^{-sT}}
    \end{align}

\bibliographystyle{aasjournalv7}
\bibliography{multiwfsbib}

\end{document}